\pdfoutput=1
\newcommand{\summary}[1]{\vspace{.1cm} \noindent \textbf{#1.}}

\documentclass[conference,compsoc]{IEEEtran}
\usepackage{tikz}
\usepackage{amsmath}

\usepackage{filecontents}
\usepackage{booktabs}
\usepackage{multirow}
\usepackage{xcolor}
\usepackage{subcaption}
\usepackage{float}
\usepackage{stfloats}
\usepackage{comment}
\usepackage{threeparttable}
\usepackage[normalem]{ulem}
\usepackage[hidelinks]{hyperref}

\usepackage{amsmath}

\usepackage{enumitem}
\setlistdepth{10}

\usepackage{threeparttable}
\usepackage{tabularx}
\usepackage{makecell}
\usepackage{soul}

\newcolumntype{L}[1]{>{\raggedright\let\newline\\\arraybackslash\hspace{0pt}}m{#1}}
\newcolumntype{C}[1]{>{\centering\let\newline\\\arraybackslash\hspace{0pt}}m{#1}}
\newcolumntype{R}[1]{>{\raggedleft\let\newline\\\arraybackslash\hspace{0pt}}m{#1}}

\ifCLASSOPTIONcompsoc
  \usepackage[nocompress]{cite}
\else
  \usepackage{cite}
\fi

\ifCLASSINFOpdf
\else
\fi

\begin{document}
\title{(Don't) Trust, but (Don't) Verify:\\
Developers' Attention to Security in AI-Generated Code}

\author{\IEEEauthorblockN{Hamza Khalid\IEEEauthorrefmark{1},
Ronald E. Thompson III\IEEEauthorrefmark{2},
Alejandra Sabater\IEEEauthorrefmark{1},\\
Perucy Mussiba\IEEEauthorrefmark{1},
Kelsey R. Fulton\IEEEauthorrefmark{3} and
Daniel Votipka\IEEEauthorrefmark{1}}
\IEEEauthorblockA{\IEEEauthorrefmark{1}Tufts University \quad \IEEEauthorrefmark{2}Philips \quad \IEEEauthorrefmark{3}Colorado School of Mines}
\IEEEauthorblockA{hamza.khalid@tufts.edu, ron.thompson\_1@philips.com, alejandra.sabater@tufts.edu,\\ perucy.mussiba@tufts.edu,
 kelsey.fulton@mines.edu, dvotipka@cs.tufts.edu}}

\maketitle

\begin{abstract}

AI coding assistants are rapidly transforming software development, but are known to produce insecure code. Prior work has measured whether AI-assisted developers produce secure code, but less is known about how they evaluate AI-generated code: whether they can identify vulnerabilities, what cues they use, and how trust shapes their decisions. This evaluation step is foundational to secure development with AI, whether using auto-complete, chat tools, or AI agents. As a first step, we conducted a remote observational study with 100 participants isolating this evaluation stage. Participants were tasked with producing secure and functional code for four C linked-list tasks. For each, participants were able to cycle through five AI-generated suggestions varying in security and functionality, select one, and edit their choice into a final submission. Participants also completed a post-study survey about their decision-making and perception of AI-generated code's security and 23 completed a more in-depth interview.

We observed no difference in selection rates between the most and least secure suggestions, suggesting participants struggled to distinguish secure from insecure code. Few performed a thorough security review; many relied on superficial cues such as visible edge-case handling blocks, and many focused on functionality or coding style. Still, those who made more edits to the suggestions produced more secure code, suggesting they could fix insecure code once they identified it. Participants' confidence in their code's security correlated with its actual security, indicating they were realistic about their ability to produce secure code with AI. Many reported distrusting the AI to produce secure code, but thought it was sufficient in this context or that it was better than they could do. Based on our results, we present recommendations to improve secure development with AI-generated code.

\end{abstract}

\IEEEpeerreviewmaketitle

\section{Introduction}
\label{sec:intro}
The recent surge in the use of artificial intelligence (AI) across domains has extended into software development, where generative AI tools are increasingly used to automate tasks like code generation and testing~\cite{stackoverflow_survey}. Tools like Claude Code and Cursor are specifically designed for integration in development, and developers report substantial productivity gains using AI~\cite{stackoverflow_survey}, which can produce large code segments from natural language prompts or automate common development workflows. In fact, Google recently reported that more than 25\% of their new code is AI generated~\cite{futureofswe}, and in the 2025 Stack Overflow developer survey, 51\% of professional developers reported using AI tools daily~\cite{stackoverflow_survey}.

This surge in AI coding tools' adoption introduces significant security concerns. %
AI-generated code can inadvertently introduce vulnerabilities into code bases, potentially at scale, and recent studies have shown that such tools often generate insecure code. A 2026 industry-scale evaluation of frontier LLMs found that roughly 45\% of generated code for 80 coding tasks still has vulnerabilities when explicit security guidelines were not included in the prompts, and while functional capability has improved sharply since 2023, security performance has remained essentially flat~\cite{veracode2026}. This result matches similar prior studies that found AI-generated code often included vulnerabilities~\cite{PearceAsleep2022,HamerAnotherCopyPaste2024}.

The presence of vulnerable code means developers must assess AI-generated code before integrating it into production environments.
Prior work has investigated whether developers are able to securely use AI-generated code at a high level, comparing the security of code produced by developers using AI and those not using AI. The results of this work have produced mixed results. Some have shown AI use may lead developers to produce insecure code~\cite{PerryInsecureAICode}; others did not observe a difference in security between participants using or not using AI~\cite{AsareCopilot,sandoval2023lostcuserstudy}, while others found developers were more likely to produce secure solutions with AI when prompted to think about security on a security-specific task (password storage)~\cite{alena-chi}. This variation in outcomes could be due to differences in task (e.g., password storage, memory management, crypto API use) or differences in AI interaction (e.g., auto-complete, interactive chat). Therefore, a more in-depth investigation of how developers actually interact with AI-generated code is needed. However, a deeper investigation is challenging, especially as developers interact more with the AI. With each interaction, the probabilistic nature of the AI's response introduces more variation, making comparisons between developers more tenuous. Additionally, as the AI does more work, this extends the length of time participants must spend in the study, potentially increasing fatigue.

Therefore, we focus first on the component foundational to the security of developers' interaction with AI, i.e., how they evaluate the security of AI-generated code. In auto-complete interactions, developers must decide whether to accept, reject, or edit an AI-generated suggestion. In chat-based interactions, developers must evaluate AI-generated code to determine how to iteratively refine prompts, request explanations, or decide the produced code is sufficient. Similarly, with semi-autonomous agents, developers must evaluate AI-generated code to iteratively provide direction and determine when the agent's work is done. If developers cannot effectively perform this evaluation step, then they will struggle with any subsequent iterative interaction with the LLM without additional interventions.

In this paper, we seek to understand how developers actually perform this foundational step of AI-generated code evaluation. By focusing on evaluation, we provide detailed insight into the prior mixed results of human-AI interaction in secure development, assessing whether developers can reliably complete manual code reviews, an often-recommended development practice~\cite{SAFECode,NISTCodeReview,microsoft_manual_review}. The answer shapes future interactive AI-assisted secure development study design: by controlling for variation in evaluation practices based on patterns identified in this work, or if developers cannot perform this manual task reliably, by changing the research focus to interventions like AI-assisted vulnerability discovery or prompting guidance.

Prior work on secure development user studies has investigated how developers evaluate code produced by other parties (e.g., other developers, development forums)~\cite{VotipkaBIBIFIQual2020,FultonBIBIFIWinter2022,fulton2024write,acar2016you, EdmundsonCodeReview, MeneelyCodeReview,PaulCodeReview,BrazCodeReview}. However, the AI context differs in two key ways: 1) developers' (dis)trust in AI may lead them to evaluate code differently and 2) AI can quickly provide many, mostly functional code suggestions. The latter reason is particularly interesting as developers do not need to review and edit the code to integrate it into their own, as they would for code from Stack Overflow. Also, they could cycle through versions to find the suggestion easiest to understand, fit their style, or are more secure, which would not be possible with code provided by another developer. These simplifications make the code easier to use and may make users aware of security issues, but also likely remove natural forcing functions for code understanding that support thorough code review.
Therefore, we specifically seek to answer the following research questions about developers' evaluation of \emph{AI-generated} code:

\smallskip
\begin{enumerate}
[label=\textbf{RQ\arabic*}]
\item Are developers able to distinguish secure AI-generated code from insecure code?
\item What process do developers follow when evaluating AI-generated code?
\item How does developers' trust in AI affect their code evaluation process?
\end{enumerate}

To address these research questions, we conducted an online observational study with 100 developers where they were instructed to produce secure and functional code in C for four linked list tasks. For each task, participants could cycle through five AI-generated code suggestions which varied in functional correctness and security, select one at a time, and edit %
the code to produce their final submission. This step of reviewing AI-generated code and deciding whether to accept, reject, or edit it represents the common interactions of code evaluation across all AI-assisted development interactions.
Our design allows us to capture two ways participants can demonstrate an ability to evaluate AI-suggested code's security (RQ1), either by selecting secure suggestions or editing the code to remove vulnerabilities.
We did not observe a difference in selection rates between the most secure and least secure AI suggestions, but when participants selected secure suggestions or modified the suggestions, their final solution contained fewer vulnerabilities.

Beyond security outcomes, we also conducted a fine-grained analysis of developers' behaviors and reasoning throughout the study to understand their development process (RQ2). To capture the data necessary for this review process, participants completed the programming tasks in a modified version of the NERDS environment~\cite{10.1145/3675741.3675750}, which enabled fine-grained logging of participant behavior, including time spent on tasks and suggestions, code edits, and whether submissions passed predefined security and functionality tests. All 100 participants completed a post-programming activity survey reflecting on their process, and 23 were randomly selected for interviews where they reviewed their actions during the programming tasks and explained their reasoning. Examining developers' evaluation processes reveals risks not visible from outcome-level analysis alone and informs the design of tools and practices to reduce passive acceptance of insecure suggestions. For example, many participants did not perform a thorough security review, but instead relied on superficial checks like the presence of explicit edge-case handling at the beginning of the suggestion, which led them to overlook vulnerabilities.

Finally, we examine developers' trust in the security of AI-generated suggestions (RQ3), as trust influences the level of scrutiny applied to code. Through survey and interview responses, we determined how AI is situated within developers' \emph{hierarchy of trust} relative to human collaborators and established resources to assess whether AI is afforded out-sized authority that may increase the risk of accepting insecure code.
We find a direct relationship between participants' security confidence and the actual security of their solution; when participants had less secure code, they were less confident. Participants had mixed perceptions about the security of AI suggestions, and those with more security experience believing that AI generally improves code security.

Because our results suggest developers struggle to manually evaluate AI-generated code, often performing limited checks and being unable to reliably identify insecure code, this indicates additional interventions are necessary to support AI code evaluation. The burden should therefore shift to the tools and the interaction mechanism itself. Based on our results, we provide recommendations for redesigning AI interaction and utilizing AI to perform security reviews or reduce the barrier to using more formal security analyses.

\section{Related Work}
\label{sec:rw}
Prior work has explored developers' use and perceptions of AI for secure development, measured the impact of AI on security outcomes, and explored developers' ability to evaluate third-party code.

\summary{Use and Perceptions of AI for Secure Development}
Prior research conducted interviews, surveys, and public artifact analyses to understand how developers report using generative AI to support security-related development tasks and how they perceive AI's utility and security in these contexts. Klemmer et al.\ interviewed 27 developers and analyzed 190 Reddit posts discussing AI's use for software development, finding developers were just beginning to use AI assistants, with limited usage recommendations from their organizations. While developers did not report trusting AI code suggestions' security, they believed it could be useful for security tasks as long as suggestions received manual review before being used~\cite{10.1145/3658644.3690283}.
Recent work has continued to investigate developer perceptions of AI assistants with a particular focus on their trust in AI suggestions' security%
~\cite{Oh24,ShaoAIToolAdoption2026,10.1145/3664646.3664757}. Oh et al.\ surveyed 238 developers regarding their reasons for using AI assistants for secure development, finding developers primarily used AI assistants because they believed they improved productivity and that they generally trusted %
AI suggestions~\cite{Oh24}. Shao et al. reported similar results in a survey of 305 students and professional developers with the professionals being more likely to %
trust the AI, often because they believed they had sufficient review processes in place to catch vulnerabilities~\cite{ShaoAIToolAdoption2026}. Finally, Brown et al.\ used %
interviews and artifact analysis to identify characteristics of AI suggestions that were more likely to be trusted and incorporated by Google developers, %
finding higher-confidence and medium-length suggestions were more likely to be accepted%
~\cite{10.1145/3664646.3664757}. %

\summary{Impact of AI on Secure Development}
Focusing on developers' security performance when using AI, Perry et al.\ had 47 participants complete development tasks with varying security implications~\cite{PerryInsecureAICode}. Comparing an AI-assisted group to a control, they found the AI users more often produced insecure code, while reporting greater confidence in their code's security. Sandoval et al.\ ran a similar study with 58 participants on C programming tasks, finding no statistically significant difference in security outcomes between AI-assisted and control groups~\cite{sandoval2023lostcuserstudy}. Asare et al.\ studied 25 participants using a counterbalanced design across simple and complex tasks and observed no significant differences in security outcomes~\cite{AsareCopilot}.
Perhaps closest to our work, Serafini et al.~\cite{alena-chi} and Oh et al.~\cite{Oh24} introduce vulnerabilities into AI-suggested code provided to 76 and 30 developers, respectively, to assess whether developers can produce secure code using a poisoned AI. Serafini et al.\ found developers produced secure code when given warnings and guidelines. However, many still accepted insecure suggestions, and reported high levels of trust in the security of vulnerable AI suggestions. Conversely, Oh et al.\ found developers struggled to produce secure code, often accepting insecure AI suggestions, especially when using a chat-based AI.

This prior work %
presents mixed results regarding developers' ability to use AI securely, though this may be due to the fact that they assess differences in outcomes, which are prone to many confounding factors. Therefore, in our work, we perform a detailed assessment of the foundational step in developers' secure use of AI-suggested code, i.e., their evaluation of suggestion security. We not only assess their outcomes, but also how they reached those outcomes through self-reported data in interviews and surveys, and detailed quantitative logs from the study environment. We also extend prior assessments of trust in AI~\cite{10.1145/3658644.3690283, Oh24, alena-chi, 10.1145/3664646.3664757} by 1) capturing behavioral expressions of trust through the suggestions developers select, the edits they make, and self-reported trust, and 2) comparing trust in AI-suggested code with the trust developers place in other common sources, like Stack Overflow and official documentation.

\summary{Evaluating third party code}
While assessing developers' evaluation of AI-generated code is novel, prior work has studied developers' use of code written by someone else, e.g., Stack Overflow or another developer. Multiple papers have assessed the impact of Stack Overflow use on developers' coding outcomes~\cite{acar2016you,VotipkaBIBIFIQual2020,FischerSOHarmful2017,OmarCodeProp}, showing developers often copy/paste vulnerabilities from insecure Stack Overflow posts. Similar to prior work in AI-assisted development, this only considers outcomes, not the evaluation process. Fulton et al.\ closed this gap by studying how 14 teams searched for vulnerabilities in each other's code during a secure programming exercise where each team was first tasked with securely implementing a program according to the same specification~\cite{FultonBIBIFIWinter2022}. They found teams often found vulnerabilities %
using functionality test cases they had developed to evaluate their own version of the program, thus benefiting from implementing the programs themselves. Later, Fulton et al.\ evaluated differences in 112 Python developers' evaluation practices when asked to \emph{write} their own secure code, \emph{read} provided code and identify vulnerabilities, or \emph{fix} vulnerabilities in provided code~\cite{fulton2024write}. Participants in the fix condition---the condition closest to our own experiment---mostly identified vulnerabilities using provided functionality test cases. %
We expand on this work by evaluating the impact of the unique aspects of AI-suggested code, which were not present in prior work. First, unlike code from Stack Overflow or other forums, AI suggestions require minimal modification to be functional, reducing the common friction of modification of suggestions to fit a codebase. %
Second, AI can provide multiple similar suggestions, unlike other developers or Fulton et al.'s experimental conditions~\cite{fulton2024write}, which avoids the friction of having to fix code that might not pass all test cases. %
Together, this reduction in friction makes it more incumbent on developers to actively conduct thorough reviews as they are less likely to need to understand the code before they can make it work, motivating our study. However, prior work has suggested developers often introduce security vulnerabilities due to a lack of awareness of the correct solution~\cite{VotipkaBIBIFIQual2020}. Because the AI can provide multiple suggestions, some of which are secure, this makes the developer aware of the secure solution when comparing suggestions enabling more secure outcomes.

\section{Methodology}
\label{sec:methods}

To examine how developers use, evaluate, and trust AI-generated code, we conducted a mixed-methods study combining behavioral observations, surveys and semi-structured interviews.
 This study was approved by our institution's ethics review board. %
 This section describes the study components, analysis, recruitment, and limitations. All study materials are available on OSF~\cite{supplemental}.

\subsection{Programming Tasks}

The study's primary component was a single online session where participants completed four linked list programming tasks in C: (1) adding an item, (2) removing an item, (3) updating an item, and (4) swapping two items. We chose linked lists in C to ensure task familiarity, given the prevalence of C and the inclusion of linked lists as a data structure in early undergraduate computer science education. %
Further, C's manual memory management and pointer arithmetic make it prone to complex security vulnerabilities (e.g., use-after-free, buffer overflows, memory leaks) that are often subtle and easily missed~\cite{RuefBIBIFI2016}. Lastly, prior work has shown that AI tools tend to introduce vulnerabilities in this context~\cite{PearceAsleep2022}. Therefore, these tasks %
provided a rich, security-critical context to evaluate developer behavior.

\subsection{AI Code Suggestions}
\label{sec:ai-suggestions}
For each task, participants were shown five AI-generated code suggestions varying in functional correctness and security. Suggestions were shown one at a time. Participants could navigate between them, select one to edit, and switch suggestions at any point before submitting a final solution. %

Suggestions were not generated in real time but were drawn from prior work which evaluated the security impact of AI-assisted programming by assessing code written by student programmers in an online experimental study~\cite{sandoval2023lostcuserstudy}. We used predetermined suggestions to ensure consistency between participants, isolating the evaluation stage---our study's focus---by controlling for differences in participants' prompting abilities or variation in AI responses.

For the five suggestions, we wanted to ensure a range of %
secure code representative of the full sample%
~\cite{sandoval2023lostcuserstudy}. To this end, we performed a security review for all 120 suggestions---30 per task---given by the AI assistants in Sandoval et al.'s paper~\cite{sandoval2023lostcuserstudy}.
We used %
security and functionality test cases provided in the prior work's replication package, and a security expert from our research team with nearly a decade of C secure programming experience manually reviewed each suggestion for additional vulnerabilities. Based on our analysis, we selected a sample of representative suggestions to cover the levels of security and functionality present in the prior study. For each task, we selected five suggestions by first binning candidates into three categories based on vulnerability count (few/none, some, many), then choosing two from the few/none and many categories and one from the some category. Within the few/none and many bins, selections were chosen to represent opposite ends of the functionality spectrum (most vs. least functional). The number of vulnerabilities for each varies by task to account for the different distributions of vulnerabilities by task in the original 120 suggestions. The \emph{Least} suggestions had 0-2 vulnerabilities depending on task, \emph{Some} suggestions had 1-2 vulnerabilities, and \emph{Most} suggestions had 2-5 vulnerabilities.

Since we used authentic AI-generated suggestions rather than constructing synthetic examples, the absolute number and type of vulnerabilities varied across tasks. For example, all the most secure suggestions had no vulnerabilities except for \emph{add} task, which had one. This was because the most secure \emph{add} task suggestion in Sandoval et al.'s dataset~\cite{sandoval2023lostcuserstudy} contained one vulnerability. We did not synthesize a vulnerability free replacement because doing so would change the distribution and style of the AI-generated code.
We therefore treated suggestion security as a task-relative property during our analysis (Section~\ref{subsec:analysis}), i.e., for each task, suggestions were categorized according to whether they contained few or no vulnerabilities, some vulnerabilities, or many vulnerabilities relative to the other suggestions for that task. This choice preserves the realism of naturally generated AI code while allowing controlled within-task comparisons.

\begin{figure}
    \centering
    \includegraphics[width=\linewidth]{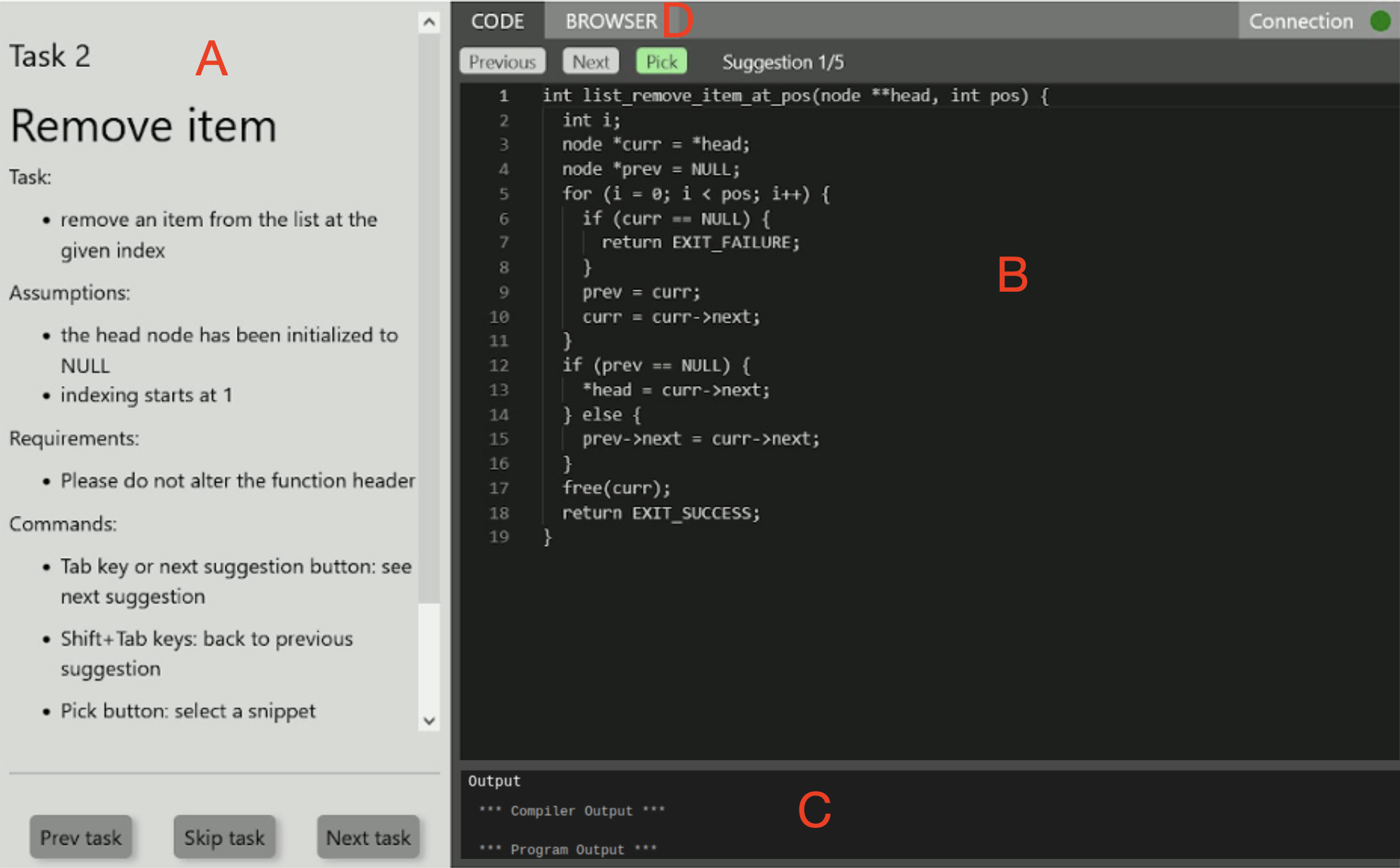}
    \caption{Screenshot of the NERDS system.}%
    \label{fig:nerds_interface}
\end{figure}

\subsection{Observational Study Data Collection}
We used a modified version of the NERDS system~\cite{10.1145/3675741.3675750} to allow participants to review the task requirements, select and edit suggestions, test their code, and submit their final solutions while capturing fine-grained interaction data. %
The NERDS system interface is shown in Figure \ref{fig:nerds_interface}.%

\summary{Task description and requirements (A)} The left side of the NERDS UI presented the current task description and requirements throughout the participants' interaction with the system. %
For example, the description for \emph{remove item} is shown in Figure~\ref{fig:nerds_interface}. %
This panel also included buttons at the bottom that allowed participants to skip the task, move to the next task, or return to the previous task. Participants were presented the tasks in a random order to avoid ordering effects.
When the participant reached the final task, the ``Next task'' button was replaced with a ``Finish'' button that redirected to the post-study survey. After pressing submit, the current state of the code for each task was saved as the participants' final submission. Until that point, participants could return to prior tasks and update their code.

\summary{Code editor (B)} The default view on the right side %
was a code editor. When participants loaded a task, they were presented with the first AI suggestion. The suggestions were presented in randomized order, and participants could scroll back and forth using the buttons at the top of the panel. Once they selected a suggestion by hitting ``Pick,'' they could begin editing the suggestion. %
However, if they later changed their mind on the best suggestion, they could return to the suggestion list and select a new one, %
but they would lose any edits made to the previously selected suggestion.

\summary{Terminal (C)} After selecting an AI suggestion, participants could execute the tests we used to assess functionality by hitting a ``Run'' button at the top of the interface, printing the results of each test to the terminal. Note, participants were not provided any test cases for security, as security tests are commonly not provided in practice~\cite{AssalDevSurvey2019,AssalSDLC18}.

\summary{Browser (D)} We gave participants access to a built-in web browser as %
part of the right panel, which we asked them to use when searching for information online, e.g., when they wanted to access external online sources like Stack Overflow. We allowed participants to visit any websites they wished as we sought to understand whether they relied on outside sources to validate the AI's suggestions~\cite{acar2016you}. Every URL visited using the built-in browser was logged.

\summary{Behavioral data collected}
The NERDS system was instrumented to log a comprehensive set of participant interactions for each task. Whenever a participant interacted with the suggestions (e.g. selected, viewed next), tested the code, or sought a resource, the system logged the time, task (add, remove, update, swap), action taken (e.g., moving between suggestions, moving between tasks, running code), browser history, terminal output, and current state of the code on the editor.
The study interface %
was specifically designed to closely mirror VS Code and GitHub Copilot, %
including allowing participants to use shortcuts to cycle through suggestions to ensure %
a realistic environment
and fine-grained data collection. Participants were provided with instructions %
detailing how the interface worked, how they could write and test their code, and how to submit their final solutions.

\summary{Post-study survey} After completing the study, participants were redirected to a survey. We first asked them about the tasks they completed (e.g., %
how they evaluated the AI suggestions, their perceived utility of the AI suggestions). %
Next, %
we asked more broadly about their experience with and perceptions of AI for software development (e.g., willingness to use AI in the future, how AI compares to other resources). The final two sections asked about programming and security experience and general demographics.

\subsection{Interviews}
To gain a deeper understanding of why participants made the choices we observed during the programming task, we conducted semi-structured interviews with a sample of 23 participants on Zoom. Audio was recorded and transcribed using Zoom's built-in services. %
Interviews lasted about 32 minutes and occurred 46 days after the session, on average.

\summary{\bf{Interview Selection}} To ensure our interviews captured a range of perspectives, we utilized a purposive sampling approach. Specifically, we sought to capture a range of task performance, i.e., from those who introduced multiple vulnerabilities to those who submitted secure code, and processes, i.e., from those who made minimal changes to the AI to those who made several changes. We bucketed participants across these two dimensions, randomly sampling from each bucket to recruit for the interview. If the selected participant chose not to respond or be interviewed, we randomly selected another from the same bucket.

\summary{Interview Protocol} We designed the interview script~\cite{supplemental} to explore participants' thought process throughout the study. The interview worked through each phase of their thought process, first asking about their process for determining which suggestion to select and what criteria they considered when assessing whether a suggestion was ``good'' or at least ``good enough.'' Then, we asked participants to discuss their reasoning behind the edits they made to the AI suggestion and, if they used outside resources, we asked them to discuss how they used them and why. We concluded each interview by asking participants to discuss their general trust in AI coding assistants, both within and beyond our study.

During the interview, we showed participants screen captures of their activity from the study. Specifically, we showed them (1) the task prompt, (2) the AI suggestion they chose for that task, and (3) their final submission to improve their recall, as it might otherwise have been difficult for them to remember the details of their actions during the study. Despite using aids, we note that recall may not have been perfect. %
This also had the benefit of allowing a more fine-grained discussion as the interviewer was able to ask questions about specific details in their process.

\subsection{Recruitment and Study Instructions}
We recruited 100 participants through three university mailing lists and Upwork for the study. A subset of 23 were selected for interviews. Participants were first directed to a pre-screening survey on Qualtrics, and had to be at least 18 years old, have experience programming in C, and be fluent in English in order to participate. The primary purpose of this screening was to ensure all participants were able to complete our tasks (which were in C), and interact with the NERDS system using English. To ensure participants were proficient in C programming, we utilized screening questions previously validated by Danilova et al.~\cite{DanilovaDoYouCode2021}. %
Only participants who correctly answered all the screening questions qualified for the study. After completing the screening survey, qualified participants were sent a link to the consent form. Once they signed the consent form, they were sent a link to our NERDS system to begin the programming tasks.

The study was advertised as an investigation of how people use AI-generated code. At the start of the study, participants were instructed to use AI-generated code suggestions to produce solutions that would be evaluated for functionality and security, and were introduced to the study environment. We did not tell participants which suggestions were secure, nor did we tell them that editing would necessarily be required to obtain a correct or secure solution.

Prior work has shown developers consider both security and functionality when security is made salient, but often do not consider security without explicit prompting~\cite{alena-chi,NaiakshinaFreelancer2019,NaiakshinaPasswordStorage2017,NaiakshinaPasswordStudents2018}. Therefore, to ensure participants attempted to produce both functional \emph{and} secure code, participants were informed that they could earn a bonus payment if their code was secure and functional. They could not receive a bonus without considering both. Participants received \$10 for completing the screening survey, an additional \$20 for completing all four study tasks, and a \$20 performance bonus for perfect scores or for performing in the top 10\% of scores in both functionality and security on their final submissions.
Ten participants qualified for the bonus. Those who completed the interviews were paid an additional \$20.

By including functionality and security incentives, we ensured we measured participants' ability to produce secure code, not just whether they considered security, while maximizing ecological validity. These incentives were designed to ensure participants were motivated to engage deeply with the tasks, and specifically with security, rather than submit code as quickly as possible. Performance bonuses have been shown to encourage participant engagement~\cite{fulton2024write}.

\subsection{Pilot}
We conducted two pilot studies to refine our methodology. First, we piloted the programming tasks observation with 20 developers. This pilot was used to test the usability of the NERDS platform, the clarity of the tasks, and the effectiveness of our incentive model. We found participants were not sufficiently motivated to thoroughly engage with the tasks without a bonus performance incentive. We therefore revised the compensation structure to add a bonus payment to better align incentives with our research goals. The pilot data was excluded from the final dataset.

For the interviews, we piloted our script with two participants to assess the clarity of our questions and the utility of presenting their own code from the task to improve recall. In the pilot, we observed that participants were hesitant to discuss external sources they used, so in the subsequent interviews, we reiterated that using external sources is completely fine, and we were only trying to understand what they did. No other major changes were required. Because the data collection process in the interview pilot and the actual interviews was almost identical and the pilot participants eventually answered---after additional probing by the interviewer---all the same questions, we chose to include our pilots' responses in our final analysis.

\subsection{Data analysis}
\label{subsec:analysis}
We used a mix of qualitative and quantitative analysis. %

\subsubsection{Qualitative analysis}
To assess the security of participants' final submissions and analyze open-ended survey responses and interviews, we utilized manual code review and iterative open coding~\cite[pg. 101-122]{strauss1998basics}, respectively.

\summary{Manual code review} Before performing quantitative analysis, we assessed participants' final code submissions' security. We reviewed %
each submission for vulnerabilities labeled in our initial code review during suggestion selection (Section~\ref{sec:ai-suggestions}). This allowed us to determine whether participants %
changed vulnerable code introduced by the AI suggestion. However, we also needed to determine if participants introduced new vulnerabilities when editing the code.

To this end, we adopted a manual security review process used in prior work~\cite{VotipkaBIBIFIQual2020,FultonBIBIFIWinter2022,fulton2024write}, in which two researchers independently reviewed each participant’s submissions for vulnerabilities. One reviewer had five years of professional and research experience, and both were provided a list of common C linked-list vulnerabilities previously identified by the researcher %
who had conducted the initial review of AI-generated suggestions used in our study (Section~\ref{sec:ai-suggestions}). This list included issues like NULL pointer dereferences, use-after-frees, missing memory releases, unchecked allocation failures, and reliance on undefined or unspecified behaviors. Reviewers expanded the list when they encountered additional vulnerabilities. The full set of unique vulnerabilities considered is provided in our supplementary materials~\cite{supplemental}. To verify a vulnerability, reviewers first identified a known vulnerable code pattern and then inspected the relevant control and data flows. For example, reviewers checked whether malformed or boundary-case inputs, such as a NULL list, an invalid position, or an allocation failure could reach the vulnerable operation and cause unsafe memory access, resource leakage, non-termination, or undefined behavior. A submission was marked as having a vulnerability if either reviewer identified it. We did not calculate agreement because these vulnerabilities were objective and verifiable; disagreements reflected only whether a reviewer had overlooked an issue, not whether it actually existed~\cite{mcdonald}.

\summary{Interviews and surveys}
Each interview transcript was manually verified and corrected where necessary. We analyzed the interviews by applying iterative open coding, identifying themes inductively~\cite[pg. 101-122]{strauss1998basics}.\footnote{Open coding is the process of uncovering, naming, and developing concepts from the data.} The interviewer and another researcher collaboratively coded two interviews to create an initial codebook.\footnote{A codebook is a set of labels (codes), each with a definition describing when it should be applied to unstructured text.} They then %
collaboratively coded the remaining interviews in rounds of four, meeting with the full research team to discuss themes and make adjustments to the codebook as needed after each round. No changes to the codebook were made after the twelfth interview.
Once all the interviews were coded, we performed axial coding to identify connections between and within %
codes~\cite[pg.123-142]{strauss1998basics}. We did not calculate inter-rater reliability (IRR)
as the interview data was primarily used to discover themes in the data and not for quantitative comparison---this is in line with an interpretivist approach, emphasizing collaborative sensemaking over statistical agreement~\cite{mcdonald}.

For open-ended survey responses, we performed iterative open coding. However, because we asked similar questions, e.g., evaluation process, security reasoning, trust in AI, we used a mix of deductive and inductive coding using the interview codebook as a starting point in our codebook development, while adding new codes as we observed additional themes. The same two coders who analyzed the interviews collaboratively coded 20 survey responses to further develop the codebook, updating and adding codes as needed. Then, %
they independently coded responses in rounds of 20, meeting after each round to calculate IRR, discuss cases where their codes differed, resolve disagreements to reach consensus, and update the codebook as necessary. We used Krippendorff's alpha ($\alpha$) to measure IRR as it accounts for chance agreements~\cite{hayes2007answering}. We repeated this process for three rounds until an $\alpha > 0.8$ was reached for all variables, which indicates acceptable reliability. One researcher coded the remaining 20 responses. We calculated IRR to allow this single-coder review due to the dataset's larger scale~\cite{mcdonald}.

\subsubsection{Quantitative analysis}
To investigate the factors influencing participants' security outcomes and decision-making when evaluating AI suggestions, we utilized a series of regressions (Table~\ref{tab:model_variables}). %

To investigate what suggestions participants selected and their final submissions' security (\emph{RQ1}), we used a Poisson regression---appropriate for count data~\cite[pg. 67-106]{cameron2013regression}---with the number of vulnerabilities in participants' final submissions as the dependent variable. We included covariates related to the suggestions' security and functionality, order the suggestion was shown in, current task, order of the task, the percentage change in lines of code %
and participants' security and programming experience. Because this regression included multiple data points from the same participant, we included a random effects variable for the participant~\cite{TUTZ1996537}.

To understand how participants evaluated AI suggestions (\emph{RQ2}), we used two regressions: 1) a logistic regression for the likelihood of participants selecting each reviewed suggestion, a binary outcome~\cite[pg. 389-405]{seltman2012experimental}, and 2) a linear regression for the time spent reviewing each suggestion, a linear outcome~\cite[pg. 213-228]{seltman2012experimental}. We included the same set of initial covariates as the previous Poisson regression, with the exception of the percentage of lines of code changed as code modifications only occur after suggestion selection and would have no impact on participants' suggestion evaluation. We also include a random effect for the participant.

Finally, to understand participants' trust in AI-generated code (\emph{RQ3}), we used four additional logistic regressions, each with binary outcomes~\cite[pg. 389-405]{seltman2012experimental}. The first regression investigated participants' reported confidence in the security of their submitted code from the post-study surveys. This was reported on a 5-point Likert-scale from ``Strongly Agree'' to ``Strongly Disagree''. We converted this to a binary variable by comparing ``Strongly Agree'' and ``Agree'' responses to the neutral and disagree options. We include the task, percentage of lines of code changed, the change in the submission's functionality score (Func. scores changed) and number of vulnerabilities (\# vulns. changed) from the selected suggestion, and the participants' security and programming experience, as initial covariates. We include a random effect for the participant because they reported confidence levels for multiple tasks.

The other three logistic regressions assessed participants' perceptions of AI coding assistants, generally. Specifically, we included regressions for participants' perception of AI's ability to produce secure code, the usefulness of AI coding assistants, and their likelihood to use an AI coding assistant in the future. For each outcome variable, we convert the 5-point Likert-scale options to a binary variable in the same way described above. For each regression, our initial covariates, which we expected could impact the outcome, include the total number of vulnerabilities across all the participant's final submissions (to allow us to compare the perceptions between participants with different security abilities), total number of external resources used, and their security and programming experience. We did not include a random effect in these regressions as they only included one response per participant.

\summary{Model fitting}
For each regression, to select a final model that was parsimonious without overfitting, we considered all possible combinations of the initial set of covariates for each model. Because the AI suggestion's security was of primary interest for our research, we only considered models that included the suggestion's security level. We also tested models including interactions for each covariate with the suggestion security. We calculated the Bayesian Information Criterion (BIC), a standard metric~\cite{raftery1995bayesian}, for each possible model and selected the minimum BIC as our final model.

\begin{table}[t]
\centering
\footnotesize
\setlength{\tabcolsep}{4pt}
\begin{tabularx}{\columnwidth}{
L{.1cm}
 p{0.23\columnwidth}
 >{\raggedright\arraybackslash}X
 c
 c
 c
}
\toprule
& \textbf{Variable} & \textbf{Definition \& Levels} & \textbf{RQ1} & \textbf{RQ2} & \textbf{RQ3} \\
\midrule
\multirow{6}{*}[-10ex]{\rotatebox[origin=c]{90}{\makecell{Outcomes}}} & \texttt{Num. vulns.} & \# of vulnerabilities present in participants' submission & X & & \\
& \texttt{Selection likelihood} & Whether a suggestion is selected & & X & \\
& \texttt{Time spent} & Log of time on suggestion & & X &  \\
& \texttt{Security confidence} & Participants' confidence in their submission's security & & & X  \\
& \texttt{AI security} & Participants' belief that AI produces secure code & & & X  \\
& \texttt{Future AI use} & Participants' plan to use AI in the future & & & X  \\
& \texttt{Usefulness of AI} & Whether participants perceive AI as useful & & & X  \\
\midrule
\multirow{10}{*}[-15ex]{\rotatebox[origin=c]{90}{\makecell{Predictors}}}&\texttt{Suggestion security} & Selected AI suggestion's security level & X & X & \\
&\texttt{Suggestion functionality} & \# of functionality unit tests passed.  & X & X & \\
&\texttt{{Submission functionality}} & {\# of func. tests passed in final submissions.}  & X & & \\
&\texttt{Suggestion order} & When was the suggestion shown (e.g., $1^{st}$, $2^{nd}$) & X & X & \\
&\texttt{Task type} & Which task (e.g., add) & X& X&  \\
&\texttt{Task order} & When was the task shown (e.g., $1^{st}$, $2^{nd}$) & X& X&  \\
&\texttt{LoC changed} & \% of lines of code changed & X &  &  \\
&\texttt{Func. scores changed} & \% change in the number of functionality tests passing & & & X \\
&\texttt{\# of vulns changed} & change in the number of vulnerabilities & &  &X  \\
&\texttt{Outside resources} & Number of resources used &  &  &X  \\
&\texttt{Security experience} & Yrs of sec exp (Binary: $\le3$ yrs, $3+$ yrs) & X & X & X \\
&\texttt{Programming experience} & Yrs of prog exp (Binary: $\le3$ yrs, $3+$ yrs) & X & X & X \\

\bottomrule\bottomrule
\end{tabularx}
\caption{Variables Used in the Regression Models: Poisson (RQ1), logistic and linear (RQ2), or logistic (RQ3).}
\label{tab:model_variables}
\end{table}

\subsection{Limitations}

As with all controlled studies of developer behavior, our design involves trade-offs between ecological validity, internal validity, and observability. Our primary goal was not to reproduce every aspect of modern AI-assisted programming, but to isolate a security-critical evaluation bottleneck: the point at which a developer must decide whether a piece of AI-generated code is safe and functional enough to adopt, edit, or reject. By controlling the exact suggestions, we ensured all participants were exposed to the same set of high-quality and low-quality suggestions, enabling robust comparison of their evaluation and selection patterns by controlling variation caused by interactive behaviors and the probabilistic nature of AI responses. While further work is necessary to assess whether our results generalize to interactive contexts where developers may refine prompts, request explanations, or ask the model to repair code, our results provide important insights into a foundational aspect of human-AI interaction for secure development and are a necessary first step to enable future investigation.

Our other ecological validity limitations match prior developer lab studies. Our tasks are relatively small and simple, and participants completed them in isolation, i.e., not with collaborators or dedicated security review. We chose this reduced scope to ensure participants were able to complete the tasks in a reasonable amount of time, to ensure tasks were solvable by developers with a range of experience, and to isolate the decision-making of individual developers. Further, we chose C as the language for our tasks because these vulnerabilities require reasoning about control and data flow, which developers struggle with~\cite{RuefBIBIFI2016}, and our design complements related work, which focused on password storage and cryptographic API use~\cite{alena-chi}. In fact, our results contradict prior work’s findings (Section 5). While this may limit generalizability to some real-world settings, our results still provide insights into AI coding assistant interaction for subtasks of a broader programming project and individual developers' initial interactions with AI suggestions.
We also chose to incentivize participants to consider security during development. A focus on security is not always common in practice~\cite{AssalDevSurvey2019,Palombo2020,kaur2025threat,fulton2024write}. However, because prior work has shown developers in lab settings will likely not consider security unless incentivized~\cite{NaiakshinaFreelancer2019,NaiakshinaPasswordStudents2018,NaiakshinaPasswordStorage2017} and the primary goal of our research was to investigate \emph{how} developers evaluate AI-suggested code's security, it was important to include this incentive to capture those behaviors. Therefore, our findings should be viewed as a likely best-case scenario of behaviors when developers are thinking about security.
For all these common design trade-offs, we chose to mirror previous similar work in human-centric secure development studies' design~\cite{PerryInsecureAICode,sandoval2023lostcuserstudy,acar2016you,Oh24,alena-chi,NaiakshinaFreelancer2019,NaiakshinaPasswordStorage2017,NaiakshinaPasswordStudents2018} to allow for general comparisons of results.

Another limitation beyond ecological validity is the representativeness of our recruited student and freelancer participants. Students often have less development experience than professionals and while freelance developers often have more experience than students, freelancers do not always have the same experience as corporate developers. However, several prior studies have shown both these populations offer adequate proxies for the broader developer community specifically in secure development~\cite{kaur2022recruit,NaiakshinaFreelancer2019,naiakshina2020conducting,tahaei2022recruiting} as many skilled professional developers lack security expertise~\cite{acar2016you,binkhorst2022security,hazhirpasand2021hurdles,mhaidli2019we,RuefBIBIFI2016,VotipkaBIBIFIQual2020}. Additionally, we follow best practice to ensure qualified developers participated using validated screening questions~\cite{DanilovaDoYouCode2021}.

Finally, to allow broad recruitment and participation by a global sample, we did not conduct in-person observations and therefore could not completely restrict participants' use of external resources. This is particularly problematic for our study because participants could choose to use an outside AI assistant to complete the task. In fact, we observed 25 participants utilize an AI assistant using our built-in browser. However, we did not observe a statistically significant difference in the amount of change between participants who used an external AI and those who did not ($U=900.5$, $p=0.771$\footnote{We used a Mann-Whitney U test to compare groups, which is an appropriate test for comparing non-parametric continuous data~\cite{mann1947test}.}) as participants using an external AI assistant changed 29 lines of code on average and those who did not changed 29.3 lines on average. This suggests these participants either used the external AI assistant to answer general questions, as they would a search engine or Q\&A website, or asked it to refine the suggestion we provided. Therefore, this does not affect our results as their responses still allow us to capture insights into their evaluation of our provided suggestions and we are able to observe their realistic use of external AI assistants during code editing.

\section{Results}
\label{sec:results}

A total of 160 people attempted our survey, 129 qualified to participate, and 105 attempted the programming tasks. %
Of these, 102 completed the tasks and the post-study survey. We removed two responses because our NERDS server experienced an unexpected network outage while they were participating which caused their logs to be corrupted and unrecoverable. This left us with 100 participants who completed the observational study, including 88 from Upwork, and 12 from universities. Among these, most were young (83\% were 18--29 years old) and identified as male (91\%). While many reported being from Pakistan (24\%), the USA (12\%), or India (12\%), the majority (52\%) were from countries that made up less than 8\% of our sample, demonstrating a geographic diversity among participants. On average, participants reported about 7 years of programming and 1.8 years of security experience.
The demographics of our sample are similar to prior developer studies~\cite{alena-chi,GeierhaasLetsHash2022,DanilovaDoYouCode2021,PerryInsecureAICode}, and are given in Table \ref{tab:demographics} in the Appendix. We conducted follow-up interviews with 23 participants with varying security and programming experience and different levels of security performance on the tasks. Demographics for each interview participant are given in Table~\ref{tab:interview-demographics}.

In total, we collected 400 task submissions (4 tasks per 100 participants),  of which 398 (99.5\%) compiled successfully. Security analysis revealed only 88 (22\%)
submissions were fully secure, i.e. free of vulnerabilities. While 42\% of participants submitted at least one secure solution, only 5\% submitted secure code for all four tasks.
The distribution of secure code varied by task, as can be seen in Table~\ref{tab:solution-security}. All secure submissions (N=88) were functional, and more functional submissions had fewer vulnerabilities ($\hat{\beta}=-0.5$, $p=0.003$; Table~\ref{tab:num-vulns}). Security rates among correct solutions by study parameter are in Appendix Table~\ref{tab:secure-frac}. %
Participants produced the most insecure code for the \emph{add} task, with 92\% of submissions being insecure and 2.9 vulnerabilities per submission on average.

Throughout the remainder of this section, we discuss which AI suggestions participants selected, how their decisions impacted their final submission, how they evaluated AI suggestions, and participants' trust in AI coding assistants. N denotes counts among all 100 participants, I among 23 interviewees, and S among 100 open-ended survey responses.

\begin{table}[t]
\centering
\footnotesize
\begin{threeparttable}
\begin{tabular}{llrrr}
\toprule
\textbf{Variable} & \textbf{Value} &\textbf{$\hat{\beta}$}  & \textbf{\textit{p}-value} & \textbf{CI} \\
\midrule
Suggestion  & Some & --\tnote{1} & --  & -- \\
Vulnerabilities & Most &  \textbf{0.4} & \textbf{\textless 0.001} & \textbf{[0.2,0.6]} \\
 & Least &  0.06 & 0.533 & [-0.1,0.2]  \\ \midrule
Change in LOC & numeric & \textbf{-0.5}  & \textbf{\textless 0.001} & \textbf{[-0.6,-0.4]} \\ \midrule
Task Type & Add & -- & -- & -- \\
 & Update & \textbf{-0.4}  & \textbf{\textless 0.001} & \textbf{[-0.6,-0.2]} \\
 & Remove & \textbf{-0.6} & \textbf{\textless 0.001} & \textbf{[-0.8,-0.4]} \\
 & Swap & \textbf{-0.2} & \textbf{0.014} & \textbf{[-0.4,-0.0]} \\ \midrule
Security       & $\le3$ years & --  & -- & -- \\
Experience & $3+$ years & -0.2 & 0.140 & [-0.4,0.0]\\ \midrule
Programming       & $\le3$ years & --  & -- & -- \\
Experience & $3+$ years & \textbf{0.2} & \textbf{0.011} & \textbf{[0.1,0.5]}\\\midrule

Submission Functionality & numeric & \textbf{-0.5}  & \textbf{0.003} & \textbf{[-0.8,-0.2]} \\

\bottomrule
\bottomrule
\end{tabular}
\begin{tablenotes}[flushleft]
        \item Statistically significant values ($p\leq0.05$) are \textbf{bolded}
        \item[1] \textbf{--} Base case (Log scale increase ($\hat{\beta}$)=0, by definition)
    \end{tablenotes}
\end{threeparttable}
\caption{Poisson regression for number of vulnerabilities in final solutions. Categorical predictors (e.g., task) are relative to the base case (first row), which has no change by definition. $\hat{\beta}$ is the log-scale increase in the outcome.}%
\label{tab:num-vulns}
\end{table}

\begin{table}[t]
\centering
    \begin{threeparttable}
    \footnotesize
    \setlength{\tabcolsep}{3pt}
    \begin{tabular}{lrcccccc}
    \toprule
          \multirow{2}{0pt}{\textbf{Factor}}  &           \textbf{Value}          & \multicolumn{3}{c}{\textbf{Likelihood}}                                      & \multicolumn{3}{c}{\textbf{$\log($Time Spent$)$}}                             \\
             &                     &\textbf{OR}              & \textbf{p-Value}         & \textbf{CI}                 & \textbf{exp}              & \textbf{p-Value}         & \textbf{CI}               \\\midrule
     Sugg. & Some  & --\tnote{1}                &        --       & --                        & --               & --               & --                      \\
           Vulns.  & Most  &   \textbf{0.6} & \textbf{\ensuremath{<}0.001} & \textbf{[0.5, 0.8]} &        1.0  &  0.491 & [0.8, 1.1]          \\
             & Least       & \textbf{0.8}  & \textbf{0.018} & \textbf{[0.6, 1.0]} & 1.0  & 0.648 & [0.8, 1.1] \\\midrule
    Task Order      & --\tnote{2}     & \textbf{1.4}                & \textbf{\ensuremath{<}0.001}             & \textbf{[1.3, 1.6]}                       & \textbf{0.9}   & \textbf{\ensuremath{<}0.001}          & \textbf{[0.9, 0.9]}                  \\\midrule
   Sugg. Index   & --\tnote{2}                  & \textbf{1.1}              & \textbf{\ensuremath{<}0.001}              & \textbf{[1.1, 1.1]}                  & \textbf{0.9}               & \textbf{\ensuremath{<}0.001}       & \textbf{[0.9, 1.0]}                    \\
                \bottomrule
             \bottomrule
    \end{tabular}
    \begin{tablenotes}[flushleft]
        \item Statistically significant values ($p\leq0.05$) are \textbf{bolded}
        \item[1] \textbf{--} Base case (OR and exp=1, by definition)
        \item[2] For linear predictors, the effect represents a 1 unit increase in the predictor.
    \end{tablenotes}
    \end{threeparttable}
    \caption{Logistic regression model for likelihood of selecting a suggestion \& linear regression model for the log of time spent evaluating a suggestion. OR is the odds ratio of increasing the outcome one unit. Exp is the exponential outcome change.}
    \label{tab:models_combo}
\end{table}

\subsection{Suggestion Selection and Code Security (RQ1)}
\label{sec:results-rq1}
To provide context for participants' decision-making, we begin by discussing the outcomes of participants' interactions with the AI suggestions. We answer two primary questions: \textbf{(1)} Which suggestions are most frequently chosen, and \textbf{(2)} How developers' suggestion choices correlate with the quality of their final submissions, measured in security (number of vulnerabilities), and functional correctness.

\summary{No observed selection difference between the most/least secure suggestions; suggestions with some vulnerabilities were selected at a higher rate} Looking first at which suggestions participants selected, we found they selected the most secure suggestions in 37.7\% of cases, suggestions with some vulnerabilities in 27.3\% of cases, and the least secure suggestions in 35.0\% of cases. Interestingly, in our logistic regression of the likelihood of selecting a particular suggestion (Table~\ref{tab:models_combo}), suggestions with some vulnerabilities were selected at a higher rate than both the least secure (OR$=0.6$, $p<0.001$) and most secure (OR$=0.8$, $p=0.018$) suggestions. Therefore, even though participants more often selected the least or most secure suggestions overall, this appears to be a factor of there being more of these suggestions (two each) than suggestions with some vulnerabilities (one), rather than participants being more likely to select those suggestions. However, this still suggests developers struggled to identify secure suggestions as these suggestions still had multiple vulnerabilities. Also, the selection estimates for the most secure and least secure suggestions had overlapping confidence intervals, indicating we did not observe a statistically significant difference. We did not observe a statistically significant difference in selection rates when controlling for final submission correctness.

\summary{Participants who selected more secure suggestions had fewer vulnerabilities in their final submissions} While we did not observe that participants were able to reliably select secure suggestions, selection security correlated with final submission security.
When participants chose more secure suggestions, i.e., a few or some vulnerabilities, they produced final submissions with 2 vulnerabilities on average, compared to 2.4 vulnerabilities on average when the least secure suggestion was chosen. This difference was statistically significant ($\hat{\beta}=0.4$, $p<0.001$) when compared to suggestions with some vulnerabilities; non-overlapping confidence intervals with least secure suggestions in Table~\ref{tab:num-vulns}. This suggests participants likely do not select insecure suggestions and fix the vulnerabilities.

\summary{\bf{More suggestion modifications resulted in fewer vulnerabilities}}
Even though the initial number of vulnerabilities in selected suggestions correlated with the number of vulnerabilities in participants' final submissions, we next wanted to know whether participants were able to fix vulnerabilities if they chose to edit the code. To evaluate the changes made to the selected suggestions, we measured the lines of code participants modified after selecting a suggestion, finding suggestions were modified by 29.2 lines on average. Our Poisson regression (Table~\ref{tab:num-vulns}) showed participants produced final submissions with statistically significantly fewer vulnerabilities the more lines they changed ($\hat{\beta}=-0.5, p < 0.001$). This indicates participants who made more changes to the code were actually able to fix some vulnerabilities.

\summary{\bf{Programming experience and task type also impacted the number of vulnerabilities}} In addition to suggestion security, which is central to our research question, we found other covariates had a statistically significant effect in our final regression model for the number of vulnerabilities in the final submissions (Table~\ref{tab:num-vulns}). Interestingly, participants who reported having more programming experience ($>3$ years) had significantly \textit{more} vulnerabilities ($\hat{\beta} = 0.2$, $p = 0.011$) in their final submissions.
Additionally, all participants who reported seeing no relevant security concerns in our tasks (Section~\ref{sec:results-rq2}) had $>3$ years of programming experience. However, the only process difference in the interviews or observations by experience was that more experienced participants spent slightly more time per suggestion on average (31 vs.\ 20 seconds).
This may suggest general programming expertise does not translate to secure coding practices, which has been observed in other secure development studies~\cite{VotipkaBIBIFIQual2020,NaiakshinaPasswordStorage2017}.
This may also indicate that when working with AI suggestions, experienced developers overestimate their ability to produce secure code. This overestimation of one's ability is a common phenomenon, especially in more complex tasks~\cite{Moore-2008-overestimation} and should be investigated further in future work.

We also observed a statistically significant effect of the task on the number of final vulnerabilities. When completing the remove, update, and swap tasks, participants had statistically significantly fewer vulnerabilities compared to the \emph{add} task (Table~\ref{tab:num-vulns}). This is likely because even the best AI-generated suggestion for the \emph{add} task included one vulnerability, so participants started from an insecure baseline even when choosing the most secure option. This provides further evidence that participants generally struggled to fix vulnerabilities in the suggestions they selected, as the distribution of vulnerabilities by task in the suggestions is reflected in the final number of vulnerabilities by task.

\subsection{Developers' Evaluation Processes (RQ2)}
\label{sec:results-rq2}

We next discuss how participants evaluated suggestions, which provides insight into the reasons for suggestion selections and submission security. From their interaction logs and survey and interview responses, we found participants typically only performed cursory security reviews and many did not conduct any form of security review for various reasons. Instead, suggestions were often chosen based on functionality or coding style. We also observed participants' review effort level appeared to decrease over time.

On average, participants spent 21 seconds per suggestion: 31 seconds for selected suggestions and 19 seconds for skipped ones. However, we noticed a gap in review time where several suggestions were reviewed in less than 2 seconds, but almost no participants reviewed suggestions for 2-5 seconds. We expect these logs with less than 2 second review times reflect participants quickly scrolling through suggestions, which they might do after they had reviewed all suggestions and want to go directly to another preferred suggestion they viewed earlier.
Because logs under two seconds likely reflect scrolling rather than evaluation, we recomputed averages excluding logs under 2 seconds, yielding a 37 seconds average overall; 43 seconds for selected suggestions, and 35 seconds for skipped suggestions. However, regressions with and without sub-two-second logs produced identical results (Section~\ref{subsubsec:RQ2-time-analysis}), so we report results for the full dataset. We found no significant effect of suggestion security or functionality on review time (Table~\ref{tab:models_combo}). %

\subsubsection{Security evaluation}
\label{sec:sec-eval}
Although participants were informed their solutions would be evaluated for security, 16 of 23 interviewees said they did not take any explicit steps to address security in their solutions, citing several reasons. Further, when participants did consider security or when they were asked how they would have assessed security for other tasks, their approaches were often cursory.

\summary{Perceived low risk led participants to dismiss security concerns}
Several interviewees dismissed the possibility of vulnerabilities in the context of the study tasks, citing the perceived low risk as justification for disregarding security considerations (I = 5). These interviewees explicitly stated that they believed the tasks posed no meaningful security risks. As P21 explained, ``The code was not something that had to be properly tested before\ldots{}passing it into the AI, so I didn’t feel there was any need to check security.''

\summary{Some lacked the knowledge to deal with security}
Other participants acknowledged their lack of review was not because they believed a security review was unimportant, but instead because their lack of security knowledge prevented them from effectively addressing security (I = 4). These interviewees explicitly stated that a lack of security knowledge was a barrier. P19 said, ``I don’t have that deep knowledge in security,'' while P8 noted they attempted to make changes only ``as much as I know, and it is very limited.''

\summary{Some participants explicitly deprioritized security} %
When asked about security considerations, some interviewees reported explicitly prioritizing functional correctness over security (I = 4). Passing unit tests served as the primary---often sole---metric of success, causing them to miss vulnerabilities. P15 noted they did not invest much effort in security because ``functionality comes first\ldots{}and security has to come second.'' Further, a subset of interviewees stated they deprioritized security, though they did not provide a specific reason for this deprioritization (I = 3). As P5 explained, ``for checking security or memory leakage\ldots{}I care least about that.'' This matches prior work which has found that developers often deprioritize security in practice in favor of functionality~\cite{AssalDevSurvey2019,Palombo2020,kaur2025threat,fulton2024write}. In fact, this number is likely small in our study only because we explicitly incentivized security~\cite{NaiakshinaFreelancer2019,NaiakshinaPasswordStudents2018,NaiakshinaPasswordStorage2017}.

\begin{figure}
    \centering
    \includegraphics[width=1\linewidth]{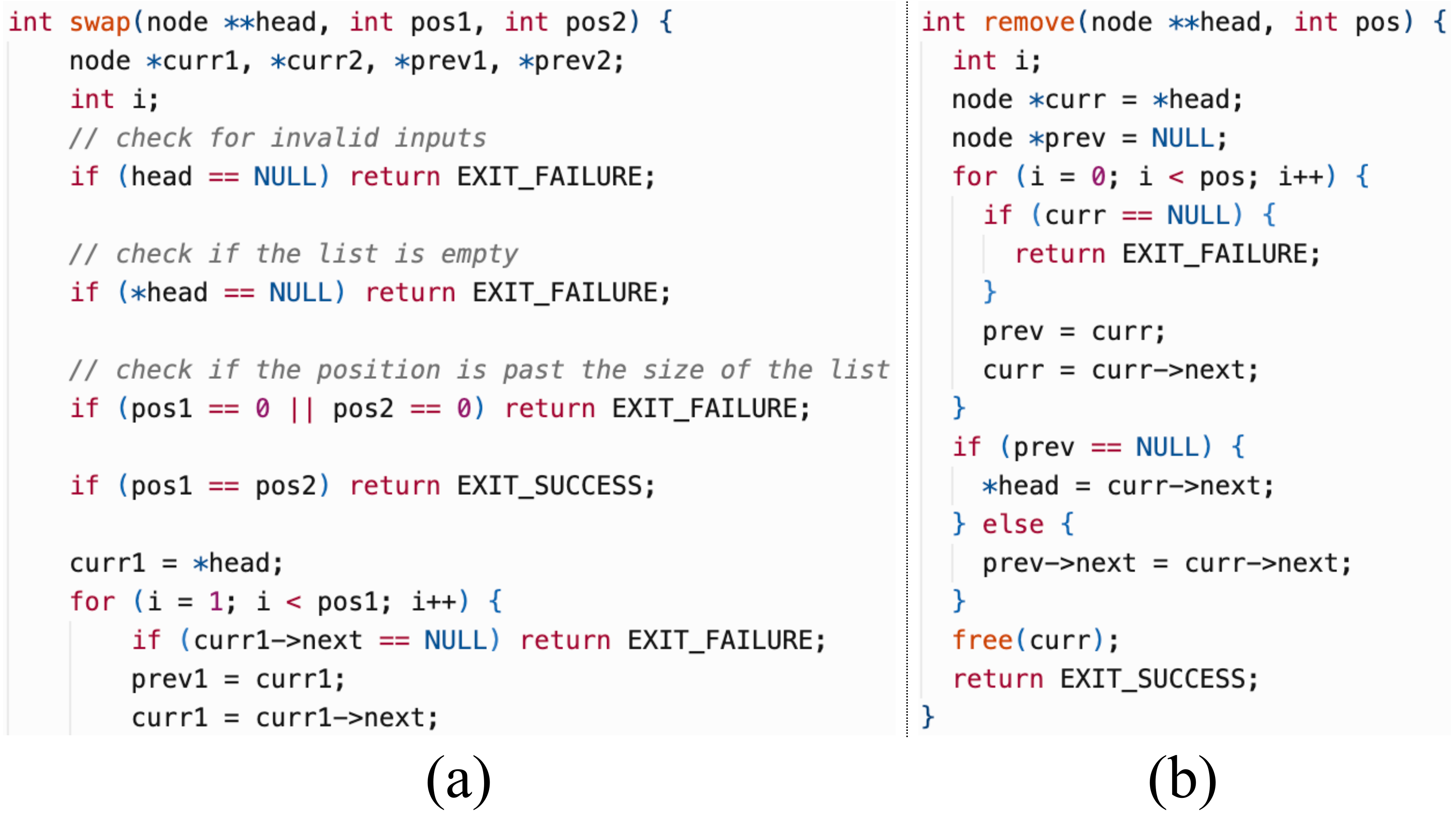}
    \caption{(a) is an insecure--no check if previous is NULL creating a circular list leaking memory--AI suggestion with a clearly marked edge case handling block. (b) is a secure AI suggestion with incorporated edge case handling.}
    \label{fig:swap_item_edge_case}
\end{figure}

\summary{Participants used the presence of edge cases as a cursory security evaluation}
While many participants indicated they did not consider security, when asked what they would have considered, the majority of interviewees (I = 18) said they assessed AI-generated suggestion security by looking at whether it included edge-case handling code. For example, P20 said they ``just look for the errors and how the code handles the errors'' when assessing a suggestion security.
Similarly, in the post-study survey, many respondents (S = 44) reported using edge-case handling to assess the security of AI-generated code. One noted they ``thought of every possible test and edge case, viewed every snippet, and chose the one implementing my tests and edge cases.''
While this type of review is simple and quick, reliance on edge-case handling alone was insufficient for assessing suggestion security. Figure \ref{fig:swap_item_edge_case} contrasts a case where a suggestion includes a clearly marked edge-case block yet exhibits poor security properties (\emph{a}), with an example with checks less visible throughout code, but no vulnerabilities (\emph{b}). This demonstrates clear edge-case handling, while salient to participants, is not a reliable standalone security indicator. Relying on a single salient surface feature in place of effortful analysis is characteristic of \emph{heuristic} rather than systematic processing, in which easily observed cues replace deeper scrutiny~\cite{Chaiken-1980-heuristic}. This mirrors AI-assisted decision-making more broadly, in which salient signals increase acceptance of AI output regardless of its correctness~\cite{bansal-2021-AI-explanations}.

\summary{Several participants recognized memory safety as an important consideration} Several interviewees discussed memory safety (I = 16) as a metric for evaluating the security of AI-generated code suggestions. These participants focused on whether code followed best practices for memory handling.
P1 emphasized the importance of proper memory deallocation, stating, ``So just make sure\ldots{}it would get freed. So as long as I did that, it was fine.'' Similarly, some post-study survey respondents (S = 17) reported using memory safety as an evaluation metric. One noted their selection process involved checking, ``Does it leak memory (forgetting to free data)?"
Our external resource usage review also showed that while only about half of participants used outside resources, memory-related queries, such as ``dynamic memory allocation in C'' were some of the most common (N = 10), only behind searches for information generally about C functions and questions about linked lists.

\summary{\bf{Participants reviewed few external resources}}
Another potential method to ensure suggestion security would be to reference external resources for known-good solutions.
While a slight majority of participants reviewed external sources (N = 53), on average, they used 3.7 resources. We count each link visited or search query on a website as a unique resource. If a participant Googled ``linked list in C'' and then ``strcpy in C'' we counted these as two unique resources. In total, participants visited 23 unique websites.

The most frequently visited domains were google.com (N = 43), chatgpt.com (N = 19), geeksforgeeks.org (N = 14), and stackoverflow.com (N = 13).
Most participants (N = 20) searched for topics related to the C language (e.g. ``strcpy in C'', ``strlen'' etc.), 18 participants made queries regarding linked lists (e.g. ``[m]odify linked list contents''), and 7 sought help with pointers (e.g. ``[p]ointers to pointers in C++''). Note, these queries did not directly reference security, though it is possible security solutions were provided by these resources, as participants may have requested security review or information. We were unable to capture these details due to the nature of our logging. We also did not observe any clear difference in editing patterns between participants who used external LLMs and those who did not.

\subsubsection{Participants considered functionality and style}
In addition to their security evaluation, participants reported a variety of other evaluation criteria when considering suggestions; the most common was a focus on code functionality (I = 22, S = 48), which usually involved running test cases. As P20 said, their evaluation involved checking if the suggestion was ``passing all the test cases or not,'' before choosing one that failed the least. Some participants (I = 4, S = 7) dry-ran each AI suggestion to trace the execution in mind before running test cases. Conversely, some selected the first suggestion, ran it to determine which functionality test cases it failed, made minor tweaks to resolve functionality errors, and repeated this until they had fully functional code (I = 10, S = 6). These participants emphasized code functionality rather than assessing how it actually worked. For instance, P21 said, when test cases failed, they knew they ``[had] to like tweak it a little bit'' to make it work.

Conversely, many participants (I = 17, S = 45) reported preferring to read and understand the suggestions in depth while evaluating them. For example, P4 said they ``went through [the suggestions] line by line, and\ldots{} worked through what [they] thought each thing did.'' This approach matches a typical code review, which opens the possibility to participants identifying vulnerabilities as they learn more about the program. Also, some participants who performed this type of review explicitly indicated they did consider security (as described in Section~\ref{sec:sec-eval}). However, most participants in this category said the reason they sought to understand the code was to ensure it was fully functional.

Participants also reported selecting suggestions that fit their personal code style preferences. Many (I = 13, S = 49) had a preference for simplicity of suggestions, prioritizing code that was easy to parse at a glance and rejecting complex solutions. P15 explained: ``I just chose a normal [for loop] here. Just because it was straightforward." Further, some participants (I = 12, S = 13) prioritized suggestions that aligned with their personal style, selecting suggestions that mimicked their own naming conventions, patterns, and general programming styles. As P1 stated, ``I'd always veer towards the one that was a little bit closer to [how] I would do things.'' Other factors like the presence of comments (I = 2, S = 2) and the code being fast (I = 1, S = 16) appeared in specific contexts in the interviews and survey responses.

\subsubsection{Exposure to tasks and suggestions}
\label{subsubsec:RQ2-time-analysis}
While considering how developers reviewed suggestions, we also wanted to explore their evaluations' changes over time.

\summary{Participants selected fewer suggestions and spent less time per suggestion with each task}
As participants worked through tasks, they selected fewer suggestions. Table~\ref{tab:models_combo} shows a significant increase in selection likelihood as tasks progressed (OR$= 1.4, p < 0.001$), with the average number of selections dropping from 2.1 for the first task, to 1.8, 1.6, and 1.5 for the second, third, and fourth tasks, respectively. This higher per-suggestion selection likelihood indicates participants reviewed fewer suggestions before settling on one as the study progressed. Participants also spent less time reviewing suggestions as they progressed. Table \ref{tab:models_combo} shows a significant reduction in time spent per suggestion ($exp= 0.9, p < 0.001$), indicating a shift toward quicker evaluation and fewer test runs. Also, the average number of test case executions decreased from 5.8 in the first to 4.5, 4.4, and 3.8 in subsequent tasks. Interview and post-study survey responses further support this observation. Several participants (I = 10, S = 3) reported changing their evaluation strategies over the course of the study. These patterns may reflect a learning effect where participants developed strong preferences for certain AI suggestions, or general fatigue, leading participants to select the first viable suggestion rather than exhaustively evaluating alternatives.

\summary{Participants were more likely to select later suggestions, but spent more time evaluating earlier suggestions} As participants reviewed more suggestions, the suggestion selection likelihood increased ($OR = 1.1, p < 0.001$). We found the second highest selection rate for the first suggestion (27\% of cases), but this drops to 12\% for the second and third suggestions and 10\% for the fourth. The fifth through tenth suggestions fluctuate between 14\% and 16\%, and the percentages slowly rise until peaking at 28\% on the seventeenth suggestion.
However, as participants reviewed more suggestions, their time spent per suggestion decreased ($exp= 0.9, p<0.001$). This suggests a learning effect
or alternatively or simultaneously, participant fatigue.

\subsection{Developer Trust in AI Suggestions (RQ3)}

Finally, we turn to developers' code security perceptions when using the AI suggestions and their views more generally of the AI-generated code's security. Their AI trust likely shapes their interactions with AI-generated code and could impact their development process beyond what we observed. %

\begin{table}[t]
\centering
\footnotesize
\begin{tabular}{lllrrr}
\toprule
\rotatebox[origin=c]{90}{\textbf{Model}} & \textbf{Factor} & \textbf{Value}  & \textbf{OR}  & \textbf{\textit{p}-value} & \textbf{CI} \\
\midrule
\multirow{2}{*}{\rotatebox[origin=c]{90}{Conf.}} & \# vulns.  & -- & \textbf{0.7} & \textbf{0.007} & \textbf{[0.55, 0.91]}  \\ \cline{2-6}\\[-0.9em]
& \# ext. resources  & -- &\textbf{0.8}  & \textbf{0.043} & \textbf{[0.7, 0.9]}  \\ \midrule

\multirow{3}{*}{\rotatebox[origin=c]{90}{AI Sec}} & sec. exp.  & $\le 3 years$   & --  \\
&  & $3+$ years & \textbf{4.9}  & \textbf{0.048} & \textbf{[1.0, 23.2]} \\ \cline{2-6}\\[-0.9em]
 & func.  & percentage & 1.0 & 0.254  & [0.9, 1.0]  \\
\bottomrule
\bottomrule
\end{tabular}
\begin{tablenotes}[flushleft]
        \item Statistically significant values ($p\leq0.05$) are \textbf{bolded}; \textbf{--} Base case
    \end{tablenotes}
\caption{Logistic regressions for participants' confidence in submission's security and belief AI makes code secure.}
\label{tab:rq3model}
\end{table}

\summary{Participants with less secure code were less confident in their code's security}
First, we consider participants' confidence in their own submissions' security to understand whether their security perceptions matched reality. %
While we did not observe differences in their ability to select secure suggestions (Section~\ref{sec:results-rq1}) and many participants reported limited security reviews (Section~\ref{sec:results-rq2}), %
we did find as participants' final submissions had more vulnerabilities, their confidence in their code's security was statistically significantly likely to decrease ($OR=0.7, p=0.007$; Table \ref{tab:rq3model}). Participants producing fewer vulnerabilities than average (N = 47) reported being confident in their submission's security 87.2\% of the time, while participants producing more vulnerabilities than average (N = 53) reported being confident in their submission's security 65.1\% of the time. While confidence is not a direct security predictor, as several participants who produced more vulnerable code still reported being confident in their submissions' security, this suggests participants who produced less secure code were at least less confident in their submission's security.

\summary{Participants who consulted more external resources were less confident}
We also observed participants who reviewed more external resources were less confident in their code's security %
(OR$=0.8, p = 0.043$) (Table \ref{tab:rq3model}).
This result is somewhat expected, as using these resources likely indicates doubts about the suggestion. As a result, their initial uncertainty may have persisted through task completion, leading to lower confidence in final submissions. We found participants often visited sites like StackOverflow even though prior work shows they also often contain security issues~\cite{acar2016you, FischerSOHarmful2017}. %
Although participants who used external resources had slightly more vulnerabilities on average (2.30) than those who did not (1.98), a Mann--Whitney U test found no significant relationship ($p=0.24$),
suggesting that consulting external materials did not reliably impact code security.%

\summary{Participants had mixed perceptions on AI suggestions' security}
61\% of participants agreed using AI generally improves code security. %
However, this perception flipped when asked if developers using AI write more secure code (%
39\% agreed). %
When comparing to other sources, such as Stack Overflow and official documentation, 25\% and 15\% of participants, respectively, believed AI suggestions would be more secure than those sources. This suggests developers believe AI might not provide the most secure code, but it might help improve code security through other means.

While many participants did not trust AI-generated code, most (89\%) intend to use AI in future, and we did not observe any statistically significant differences in expected future use between participants of varying backgrounds (Table~\ref{tab:future-AI-use-regression}). %
Most agreed AI suggestions are more useful than Stack Overflow (61\%) and official documentation (36\%).

Taken together with the security-specific skepticism above, this preference for AI over human-generated code for usefulness, but not security, reflects \emph{algorithm appreciation}---the tendency to weigh algorithmic advice more heavily than human advice~\cite{Logg-2019-appreciation}---operates selectively: participants appreciated AI for its usefulness while remaining averse for the harder-to-verify security aspect. This matches prior work showing algorithmic advice tends to be trusted more for objective rather than subjective tasks~\cite{Castelo-2019-aversion}. %

\summary{Participants with more security experience were more likely to believe AI improves code security}
We observed a slight difference in trust when comparing participants with different levels of security experience in our regression (Table \ref{tab:rq3model}). Participants with $>3$ years of security experience were significantly more likely to report confidence in AI's ability to increase code security ($OR = 4.9, p=0.048$). Among participants with $>3$ years of security experience, the vast majority (85.7\%) believed AI tools generally increase security, with only 14.3\% expressing skepticism. In contrast, participants with $\le3$ years of experience were more divided: only 57.0\% believed AI increases security, while 43.0\% felt it did not. This data suggests experienced security practitioners are more optimistic about the potential for AI to enhance code security. This is to some extent surprising, though it matches some prior work~\cite{ShaoAIToolAdoption2026}.

\section{Discussion and Conclusion}
\label{sec:discussion}

Our results indicate developers struggled to evaluate AI-generated code, often selecting insecure suggestions or failing to fix them, leaving vulnerabilities in final submissions. Evaluations were often cursory scans for security signals rather than thorough code reviews. However, participants who edited suggestions were less likely to submit vulnerabilities, suggesting they could often fix identified vulnerabilities. Participants with less secure submissions also reported lower confidence in their security. %

These findings help explain prior work showing developers produced insecure code when using AI systems poisoned to generate vulnerabilities~\cite{Oh24}, yet found and fixed AI-generated vulnerabilities after an intervention~\cite{alena-chi}. This difference may reflect task characteristics: password storage relies on well-known security patterns, %
whereas finding C memory-safety vulnerabilities requires reasoning about control and data flow, a task developers often struggle with~\cite{RuefBIBIFI2016}. %

\summary{Design interactions to support structured, thorough review} We provided participants full code solutions as AI suggestions. Because participants often only performed quick, cursory reviews, the larger code suggestion may be detrimental to their process, potentially causing developers to be overwhelmed with code, where they might otherwise be able to fix vulnerabilities if they took the time to understand the suggestion. Interactions should be changed to encourage a structured review.
This limited, structural approach has  shown promise for other tasks~\cite{yang2025overload,DangCorpusStudio2025,CollomosseAuthenticity2024,robe2022designing,yen2023coladder,ZhouScaffolding2023,AngertSpellburst2023}. Future work could explore similar approaches, such as limiting the AI suggestion's scope, e.g., to one function or line of code, to see whether this encourages more thoughtful interaction. Alternatively, AI could indicate where suggested code was drawn from and if it was taken from a secure source (e.g., a codebase previously validated by security experts), limiting the scope of required analysis.

\summary{Make use of alternative interaction paradigms to reduce developer security review} Because participants struggled to identify vulnerabilities, interaction designs should reduce developer security review. One approach is specification-driven interaction, where the AI produces a specification of what the code \emph{should} do, which the developer edits to restrict AI implementation. Cursor’s Plan Mode~\cite{CursorPlanMode} and Amazon’s Kiro~\cite{AmazonKiro} employ this, and developers would benefit from integrating these tools into their workflows.

\summary{Use AI to find vulnerabilities and support program analysis and testing tools}
Asking developers to perform security reviews of AI-generated code may ultimately be insufficient. Participants appeared aware of their security limitations, suggesting they might be open to a new approach.
One option with recent success~\cite{project_glasswing} is to ask AI to search for vulnerabilities and have developers triage results. Also, AI could produce a harness for traditional security analyses, e.g., fuzzing~\cite{bounimova2013billions,ossfuzz} or property-based testing~\cite{FinkPBT1997}. These offer more robust security guarantees, but
are often unused as they are hard to set up~\cite{Ploger2021,PlogerAFLLibFuzz2023,zhao2025qualitative}. This shifts developer interaction to providing property descriptions or directing fuzzers, which are tasks to which they may be better suited.

\section*{Acknowledgments}
We thank the study participants and interviewees for their participation. We thank the anonymous reviewers and the shepherd who provided helpful comments and constructive feedback on this paper. This project was supported by a gift from Cisco and NSF grant CNS-2440353.

\bibliographystyle{plain}
\bibliography{references}

@inproceedings{zhao2025qualitative,
author = {Zhao, Yunze and Guo, Wentao and Goldstein, Harrison and Votipka, Daniel and Fulton, Kelsey R. and Mazurek, Michelle L.},
title = {{A Qualitative Analysis of Fuzzer Usability and Challenges}},
year = {2025},
isbn = {9798400715259},
publisher = {Association for Computing Machinery},
address = {New York, NY, USA},
url = {https://doi.org/10.1145/3719027.3765055},
doi = {10.1145/3719027.3765055},
booktitle = {Proceedings of the 2025 ACM SIGSAC Conference on Computer and Communications Security},
pages = {2504–2518},
numpages = {15},
location = {Taipei, Taiwan},
series = {CCS '25}
}

@INPROCEEDINGS{DanilovaDoYouCode2021,
  author={Danilova, Anastasia and Naiakshina, Alena and Horstmann, Stefan and Smith, Matthew},
  booktitle = {Proceedings of the 43rd International Conference on Software Engineering},
  title={{Do you Really Code? Designing and Evaluating Screening Questions for Online Surveys with Programmers}}, 
  year={2021},
  volume={},
  number={},
  pages={537-548},
  doi={10.1109/ICSE43902.2021.00057},
  location = {Madrid, Spain},
  series = {ICSE '21}}

@article{raftery1995bayesian,
 ISSN = {00811750, 14679531},
 URL = {http://www.jstor.org/stable/271063},
 author = {Adrian E. Raftery},
 journal = {Sociological Methodology},
 pages = {111--163},
 publisher = {[American Sociological Association, Wiley, Sage Publications, Inc.]},
 title = {{Bayesian Model Selection in Social Research}},
 urldate = {2026-06-11},
 volume = {25},
 year = {1995}
}

@inproceedings{DangCorpusStudio2025,
author = {Dang, Hai and Swoopes, Chelse and Buschek, Daniel and Glassman, Elena L.},
title = {{CorpusStudio: Surfacing Emergent Patterns In A Corpus Of Prior Work While Writing}},
year = {2025},
isbn = {9798400713941},
publisher = {Association for Computing Machinery},
address = {New York, NY, USA},
url = {https://doi.org/10.1145/3706598.3713974},
doi = {10.1145/3706598.3713974},
booktitle = {Proceedings of the 2025 CHI Conference on Human Factors in Computing Systems},
articleno = {1211},
numpages = {19},
location = {
},
series = {CHI '25}
}

@article{CollomosseAuthenticity2024,
author = {Collomosse, John and Parsons, Andy and Potel, Mike},
title = {{To Authenticity, and Beyond! Building Safe and Fair Generative AI Upon the Three Pillars of Provenance}},
year = {2024},
issue_date = {May-June 2024},
publisher = {IEEE Computer Society Press},
address = {Washington, DC, USA},
volume = {44},
number = {3},
issn = {0272-1716},
url = {https://doi.org/10.1109/MCG.2024.3380168},
doi = {10.1109/MCG.2024.3380168},
journal = {IEEE Comput. Graph. Appl.},
month = may,
pages = {82–90},
numpages = {9}
}

@inproceedings{bounimova2013billions,
author = {Bounimova, Ella and Godefroid, Patrice and Molnar, David},
title = {{Billions and billions of constraints: Whitebox fuzz testing in production}},
year = {2013},
isbn = {9781467330763},
publisher = {IEEE Press},
booktitle = {Proceedings of the 2013 International Conference on Software Engineering},
pages = {122–131},
numpages = {10},
location = {San Francisco, CA, USA},
series = {ICSE '13}
}

@article{FinkPBT1997,
author = {Fink, George and Bishop, Matt},
title = {{Property-based testing: a new approach to testing for assurance}},
year = {1997},
issue_date = {July 1997},
publisher = {Association for Computing Machinery},
address = {New York, NY, USA},
volume = {22},
number = {4},
issn = {0163-5948},
url = {https://doi.org/10.1145/263244.263267},
doi = {10.1145/263244.263267},
journal = {SIGSOFT Softw. Eng. Notes},
month = jul,
pages = {74–80},
numpages = {7}
}

@inproceedings {Ploger2021,
author = {Stephan Pl{\"o}ger and Mischa Meier and Matthew Smith},
title = {{A Qualitative Usability Evaluation of the Clang Static Analyzer and {libFuzzer} with {CS} Students and {CTF} Players}},
booktitle = {Seventeenth Symposium on Usable Privacy and Security (SOUPS 2021)},
year = {2021},
isbn = {978-1-939133-25-0},
pages = {553--572},
url = {https://www.usenix.org/conference/soups2021/presentation/ploger},
publisher = {USENIX Association},
month = aug
}

@inproceedings{PlogerAFLLibFuzz2023,
author = {Pl\"{o}ger, Stephan and Meier, Mischa and Smith, Matthew},
title = {{A Usability Evaluation of AFL and libFuzzer with CS Students}},
year = {2023},
isbn = {9781450394215},
publisher = {Association for Computing Machinery},
address = {New York, NY, USA},
url = {https://doi.org/10.1145/3544548.3581178},
doi = {10.1145/3544548.3581178},
booktitle = {Proceedings of the 2023 CHI Conference on Human Factors in Computing Systems},
articleno = {186},
numpages = {18},
location = {Hamburg, Germany},
series = {CHI '23}
}

@inproceedings{ZhouScaffolding2023,
author = {Zhou, Haoquan and Li, Jingbo},
title = {{A Case Study on Scaffolding Exploratory Data Analysis for AI Pair Programmers}},
year = {2023},
isbn = {9781450394222},
publisher = {Association for Computing Machinery},
address = {New York, NY, USA},
url = {https://doi.org/10.1145/3544549.3583943},
doi = {10.1145/3544549.3583943},
booktitle = {Extended Abstracts of the 2023 CHI Conference on Human Factors in Computing Systems},
articleno = {561},
numpages = {7},
location = {Hamburg, Germany},
series = {CHI EA '23}
}

@inproceedings{AngertSpellburst2023,
author = {Angert, Tyler and Suzara, Miroslav and Han, Jenny and Pondoc, Christopher and Subramonyam, Hariharan},
title = {{Spellburst: A Node-based Interface for Exploratory Creative Coding with Natural Language Prompts}},
year = {2023},
isbn = {9798400701320},
publisher = {Association for Computing Machinery},
address = {New York, NY, USA},
url = {https://doi.org/10.1145/3586183.3606719},
doi = {10.1145/3586183.3606719},
booktitle = {Proceedings of the 36th Annual ACM Symposium on User Interface Software and Technology},
articleno = {100},
numpages = {22},
location = {San Francisco, CA, USA},
series = {UIST '23}
}

@book{seltman2012experimental,
  title={{Experimental design and analysis}},
  author={Seltman, Howard J},
  year={2012},
  publisher={Carnegie Mellon University Pittsburgh}
}

@inproceedings{yen2023coladder,
author = {Yen, Ryan and Zhu, Jiawen Stefanie and Suh, Sangho and Xia, Haijun and Zhao, Jian},
title = {{CoLadder: Manipulating Code Generation via Multi-Level Blocks}},
year = {2024},
isbn = {9798400706288},
publisher = {Association for Computing Machinery},
address = {New York, NY, USA},
url = {https://doi.org/10.1145/3654777.3676357},
doi = {10.1145/3654777.3676357},
booktitle = {Proceedings of the 37th Annual ACM Symposium on User Interface Software and Technology},
articleno = {11},
numpages = {20},
location = {Pittsburgh, PA, USA},
series = {UIST '24}
}

@article{robe2022designing,
author = {Robe, Peter and Kuttal, Sandeep Kaur},
title = {{Designing PairBuddy—A Conversational Agent for Pair Programming}},
year = {2022},
issue_date = {August 2022},
publisher = {Association for Computing Machinery},
address = {New York, NY, USA},
volume = {29},
number = {4},
issn = {1073-0516},
url = {https://doi.org/10.1145/3498326},
doi = {10.1145/3498326},
journal = {ACM Transactions on Computer-Human Interaction (TOCHI)},
month = may,
articleno = {34},
numpages = {44}
}

@conference {ossfuzz,
author = {Kostya Serebryany},
title = {{{OSS-Fuzz} - Google{\textquoteright}s continuous fuzzing service for open source software}},
booktitle = {Proceedings of the 2017 USENIX Security Symposium},
year = {2017},
address = {Vancouver, BC},
publisher = {USENIX Association},
month = aug
}

@article{yang2025overload,
  title={{From Overload to Insight: Scaffolding Creative Ideation through Structuring Inspiration}},
  author={Yang, Yaqing and Mohanty, Vikram and Martelaro, Nikolas and Kittur, Aniket and Chen, Yan-Ying and Hong, Matthew K},
  journal={arXiv preprint arXiv:2504.15482},
  year={2025},
  url={https://arxiv.org/abs/2504.15482}
}

@article{mann1947test,
 ISSN = {00034851},
 URL = {http://www.jstor.org/stable/2236101},
 author = {H. B. Mann and D. R. Whitney},
 journal = {The Annals of Mathematical Statistics},
 number = {1},
 pages = {50--60},
 publisher = {Institute of Mathematical Statistics},
 title = {{On a Test of Whether one of Two Random Variables is Stochastically Larger than the Other}},
 volume = {18},
 year = {1947}
}

@book{strauss1998basics,
  title={{Basics of Qualitative Research: Techniques and Procedures for Developing Grounded Theory}},
  author={Strauss, Anselm and Corbin, Juliet},
  year={1998},
  publisher={Sage Publications, Inc.}
}

@article{hayes2007answering,
author = {Andrew F. Hayes and Klaus Krippendorff},
title = {{Answering the Call for a Standard Reliability Measure for Coding Data}},
journal = {Communication Methods and Measures},
volume = {1},
number = {1},
pages = {77--89},
year = {2007},
publisher = {Routledge},
doi = {10.1080/19312450709336664},
URL = {https://doi.org/10.1080/19312450709336664},
eprint = {https://doi.org/10.1080/19312450709336664}
}

@INPROCEEDINGS{PearceAsleep2022,
  author={Pearce, Hammond and Ahmad, Baleegh and Tan, Benjamin and Dolan-Gavitt, Brendan and Karri, Ramesh},
  booktitle={2022 IEEE Symposium on Security and Privacy (SP)}, 
  title={{Asleep at the Keyboard? Assessing the Security of GitHub Copilot’s Code Contributions}}, 
  year={2022},
  volume={},
  number={},
  pages={754-768},
  doi={10.1109/SP46214.2022.9833571}}

@inproceedings{RuefBIBIFI2016,
author = {Ruef, Andrew and Hicks, Michael and Parker, James and Levin, Dave and Mazurek, Michelle L. and Mardziel, Piotr},
title = {{Build It, Break It, Fix It: Contesting Secure Development}},
year = {2016},
isbn = {9781450341394},
publisher = {Association for Computing Machinery},
address = {New York, NY, USA},
url = {https://doi.org/10.1145/2976749.2978382},
doi = {10.1145/2976749.2978382},
booktitle = {Proceedings of the 2016 ACM SIGSAC Conference on Computer and Communications Security},
pages = {690–703},
numpages = {14},
location = {Vienna, Austria},
series = {CCS '16}
}

@inproceedings{10.1145/3675741.3675750,
author = {Lewis, Joseph and Fulton, Kelsey R.},
title = {{NERDS: A Non-invasive Environment for Remote Developer Studies}},
year = {2024},
isbn = {9798400709579},
publisher = {Association for Computing Machinery},
address = {New York, NY, USA},
url = {https://doi.org/10.1145/3675741.3675750},
doi = {10.1145/3675741.3675750},
booktitle = {Proceedings of the 17th Cyber Security Experimentation and Test Workshop},
pages = {74–82},
numpages = {9},
location = {Philadelphia, PA, USA},
series = {CSET '24}
}

@INPROCEEDINGS{HamerAnotherCopyPaste2024,
  author={Hamer, Sivana and d’Amorim, Marcelo and Williams, Laurie},
  booktitle={2024 IEEE Security and Privacy Workshops (SPW)}, 
  title={{Just another copy and paste? Comparing the security vulnerabilities of ChatGPT generated code and StackOverflow answers}}, 
  year={2024},
  volume={},
  number={},
  pages={87-94},
  doi={10.1109/SPW63631.2024.00014},
  url = {https://doi.ieeecomputersociety.org/10.1109/SPW63631.2024.00014},
   publisher = {IEEE Computer Society},
   address = {Los Alamitos, CA, USA},
   month =May}

@inproceedings {VotipkaBIBIFIQual2020,
author = {Daniel Votipka and Kelsey R. Fulton and James Parker and Matthew Hou and Michelle L. Mazurek and Michael Hicks},
title = {{Understanding security mistakes developers make: Qualitative analysis from Build It, Break It, Fix It}},
booktitle = {29th USENIX Security Symposium (USENIX Security 20)},
year = {2020},
isbn = {978-1-939133-17-5},
pages = {109--126},
url = {https://www.usenix.org/conference/usenixsecurity20/presentation/votipka-understanding},
publisher = {USENIX Association},
month = aug
}

@inproceedings{PerryInsecureAICode,
author = {Perry, Neil and Srivastava, Megha and Kumar, Deepak and Boneh, Dan},
title = {{Do Users Write More Insecure Code with AI Assistants?}},
year = {2023},
isbn = {9798400700507},
publisher = {Association for Computing Machinery},
address = {New York, NY, USA},
url = {https://doi.org/10.1145/3576915.3623157},
doi = {10.1145/3576915.3623157},
booktitle = {Proceedings of the 2023 ACM SIGSAC Conference on Computer and Communications Security},
pages = {2785–2799},
numpages = {15},
location = {Copenhagen, Denmark},
series = {CCS '23}
}

@inproceedings{NaiakshinaFreelancer2019,
author = {Naiakshina, Alena and Danilova, Anastasia and Gerlitz, Eva and von Zezschwitz, Emanuel and Smith, Matthew},
title = {{"If you want, I can store the encrypted password": A Password-Storage Field Study with Freelance Developers}},
year = {2019},
isbn = {9781450359702},
publisher = {Association for Computing Machinery},
address = {New York, NY, USA},
url = {https://doi.org/10.1145/3290605.3300370},
doi = {10.1145/3290605.3300370},booktitle = {Proceedings of the 2019 CHI Conference on Human Factors in Computing Systems},
pages = {1–12},
numpages = {12},
location = {Glasgow, Scotland Uk},
series = {CHI '19}
}

@article{ShaoAIToolAdoption2026,
title = {{Empirical analysis of generative AI tool adoption in software development}},
journal = {Information and Software Technology},
volume = {192},
pages = {108036},
year = {2026},
issn = {0950-5849},
doi = {https://doi.org/10.1016/j.infsof.2026.108036},
url = {https://www.sciencedirect.com/science/article/pii/S095058492600025X},
author = {Deo Shao and Fredrick Ishengoma}
}

@inproceedings {kaur2025threat,
author = {Harjot Kaur and Carson Powers and Ronald E. Thompson III and Sascha Fahl and Daniel Votipka},
title = {{"Threat modeling is very formal, it{\textquoteright}s very technical, and also very hard to do correctly": Investigating Threat Modeling Practices in {Open-Source} Software Projects}},
booktitle = {34th USENIX Security Symposium (USENIX Security 25)},
year = {2025},
isbn = {978-1-939133-52-6},
address = {Seattle, WA},
pages = {2125--2144},
url = {https://www.usenix.org/conference/usenixsecurity25/presentation/kaur},
publisher = {USENIX Association},
month = aug
}

@inproceedings{tahaei2022recruiting,
author = {Tahaei, Mohammad and Vaniea, Kami},
title = {{Recruiting Participants With Programming Skills: A Comparison of Four Crowdsourcing Platforms and a CS Student Mailing List}},
year = {2022},
isbn = {9781450391573},
publisher = {Association for Computing Machinery},
address = {New York, NY, USA},
url = {https://doi.org/10.1145/3491102.3501957},
doi = {10.1145/3491102.3501957},
booktitle = {Proceedings of the 2022 CHI Conference on Human Factors in Computing Systems},
articleno = {590},
numpages = {15},
location = {New Orleans, LA, USA},
series = {CHI '22}
}

@inproceedings{10.1145/3658644.3690283,
author = {Klemmer, Jan H. and Horstmann, Stefan Albert and Patnaik, Nikhil and Ludden, Cordelia and Burton, Cordell and Powers, Carson and Massacci, Fabio and Rahman, Akond and Votipka, Daniel and Lipford, Heather Richter and Rashid, Awais and Naiakshina, Alena and Fahl, Sascha},
title = {{Using AI Assistants in Software Development: A Qualitative Study on Security Practices and Concerns}},
year = {2024},
isbn = {9798400706363},
publisher = {Association for Computing Machinery},
address = {New York, NY, USA},
url = {https://doi.org/10.1145/3658644.3690283},
doi = {10.1145/3658644.3690283},
booktitle = {Proceedings of the 2024 on ACM SIGSAC Conference on Computer and Communications Security},
pages = {2726–2740},
numpages = {15},
location = {Salt Lake City, UT, USA},
series = {CCS '24}
}

@INPROCEEDINGS{Oh24,
  author={Oh, Sanghak and Lee, Kiho and Park, Seonhye and Kim, Doowon and Kim, Hyoungshick},
  booktitle={2024 IEEE Symposium on Security and Privacy (SP)}, 
  title={{Poisoned ChatGPT Finds Work for Idle Hands: Exploring Developers’ Coding Practices with Insecure Suggestions from Poisoned AI Models}}, 
  year={2024},
  volume={},
  number={},
  pages={1141-1159},
  doi={10.1109/SP54263.2024.00046}}

@inproceedings{sandoval2023lostcuserstudy,
author = {Sandoval, Gustavo and Pearce, Hammond and Nys, Teo and Karri, Ramesh and Garg, Siddharth and Dolan-Gavitt, Brendan},
title = {{Lost at C: a user study on the security implications of large language model code assistants}},
year = {2023},
isbn = {978-1-939133-37-3},
publisher = {USENIX Association},
address = {USA},
booktitle = {Proceedings of the 32nd USENIX Conference on Security Symposium},
articleno = {124},
numpages = {18},
location = {Anaheim, CA, USA},
series = {SEC '23}
}

@misc{futureofswe, 
    title={{AI Writes Over 25\% Of Code At Google—What Does The Future Look Like For Software Engineers?}}, 
    url={https://www.forbes.com/sites/jackkelly/2024/11/01/ai-code-and-the-future-of-software-engineers/}, 
    journal={Forbes}, 
    author={Kelly, Jack}, 
    year={2024}
}

@inproceedings{10.1145/3664646.3664757,
author = {Brown, Adam and D'Angelo, Sarah and Murillo, Ambar and Jaspan, Ciera and Green, Collin},
title = {{Identifying the Factors That Influence Trust in AI Code Completion}},
year = {2024},
isbn = {9798400706851},
publisher = {Association for Computing Machinery},
address = {New York, NY, USA},
url = {https://doi.org/10.1145/3664646.3664757},
doi = {10.1145/3664646.3664757},
booktitle = {Proceedings of the 1st ACM International Conference on AI-Powered Software},
pages = {1–9},
numpages = {9},
location = {Porto de Galinhas, Brazil},
series = {AIware 2024}
}

@inproceedings {kaur2022recruit,
author = {Harjot Kaur and Sabrina Klivan and Daniel Votipka and Yasemin Acar and Sascha Fahl},
title = {{Where to Recruit for Security Development Studies: Comparing Six Software Developer Samples}},
booktitle = {31st USENIX Security Symposium (USENIX Security 22)},
year = {2022},
isbn = {978-1-939133-31-1},
address = {Boston, MA},
pages = {4041--4058},
url = {https://www.usenix.org/conference/usenixsecurity22/presentation/kaur},
publisher = {USENIX Association},
month = aug
}

@inproceedings{AsareCopilot,
author = {Asare, Owura and Nagappan, Meiyappan and Asokan, N.},
title = {{A User-centered Security Evaluation of Copilot}},
year = {2024},
isbn = {9798400702174},
publisher = {Association for Computing Machinery},
address = {New York, NY, USA},
url = {https://doi.org/10.1145/3597503.3639154},
doi = {10.1145/3597503.3639154},
booktitle = {Proceedings of the IEEE/ACM 46th International Conference on Software Engineering},
articleno = {158},
numpages = {11},
location = {Lisbon, Portugal},
series = {ICSE '24}
}

@inproceedings {GeierhaasLetsHash2022,
author = {Lisa Geierhaas and Anna-Marie Ortloff and Matthew Smith and Alena Naiakshina},
title = {{{Let{\textquoteright}s} Hash: Helping Developers with Password Security}},
booktitle = {Eighteenth Symposium on Usable Privacy and Security (SOUPS 2022)},
year = {2022},
isbn = {978-1-939133-30-4},
address = {Boston, MA},
pages = {503--522},
url = {https://www.usenix.org/conference/soups2022/presentation/geierhaas},
publisher = {USENIX Association},
month = aug
}

@inproceedings{alena-chi,
author = {Serafini, Raphael and Yardim, Asli and Naiakshina, Alena},
title = {{Exploring the Impact of Intervention Methods on Developers’ Security Behavior in a Manipulated ChatGPT Study}},
year = {2025},
isbn = {9798400713941},
publisher = {Association for Computing Machinery},
address = {New York, NY, USA},
url = {https://doi.org/10.1145/3706598.3713989},
doi = {10.1145/3706598.3713989},
booktitle = {Proceedings of the 2025 CHI Conference on Human Factors in Computing Systems},
articleno = {520},
numpages = {26},
location = {
},
series = {CHI '25}
}

@article{TUTZ1996537,
author = {Tutz, Gerhard and Hennevogl, Wolfgang},
title = {{Random Effects in Ordinal Regression Models}},
year = {1996},
issue_date = {Sept. 30, 1996},
publisher = {Elsevier Science Publishers B. V.},
address = {NLD},
volume = {22},
number = {5},
issn = {0167-9473},
url = {https://doi.org/10.1016/0167-9473(96)00004-7},
doi = {10.1016/0167-9473(96)00004-7},
journal = {Computational Statistics \& Data Analysis},
month = sep,
pages = {537–557},
numpages = {21}
}

@book{cameron2013regression, place={Cambridge}, edition={2}, series={Econometric Society Monographs}, title={Regression Analysis of Count Data}, publisher={Cambridge University Press}, author={Cameron, A. Colin and Trivedi, Pravin K.}, year={2013}, collection={Econometric Society Monographs}}

@article{mcdonald,
author = {McDonald, Nora and Schoenebeck, Sarita and Forte, Andrea},
title = {{Reliability and Inter-rater Reliability in Qualitative Research: Norms and Guidelines for CSCW and HCI Practice}},
year = {2019},
issue_date = {November 2019},
publisher = {Association for Computing Machinery},
address = {New York, NY, USA},
number = {CSCW},
url = {https://doi.org/10.1145/3359174},
doi = {10.1145/3359174},
journal = {Proceedings of the ACM on Human-Computer Interaction},
month = nov,
articleno = {72},
numpages = {23}
}

@inproceedings {fulton2024write,
author = {Kelsey R. Fulton and Joseph Lewis and Nathan Malkin and Michelle L. Mazurek},
title = {{Write, Read, or Fix? Exploring Alternative Methods for Secure Development Studies}},
booktitle = {Twentieth Symposium on Usable Privacy and Security (SOUPS 2024)},
year = {2024},
isbn = {978-1-939133-42-7},
address = {Philadelphia, PA},
pages = {81--100},
url = {https://www.usenix.org/conference/soups2024/presentation/fulton},
publisher = {USENIX Association},
month = aug
}

@misc{NISTCodeReview,
title={{Guidelines on Minimum Standards for
Developer Verification of Software}},
howpublished={\url{https://nvlpubs.nist.gov/nistpubs/ir/2021/NIST.IR.8397.pdf}},
publisher={National Institute of Standards and Technology},
author={National Institute of Standards and Technology},
year={2021},
month={October}
}

@techreport{SAFECode,
    title={{Fundamental Practices for Secure Software Development}},
    institution={Software Assurance Forum for Excellence in Code},
    author={Tony Rice and Josh Brown-White and Tania Skinner and Nick Ozmore and Nazira Carlage and Wendy Poland and Eric Heitzman and Danny Dhillon},
    howpublished = {\url{https://safecode.org/wp-content/uploads/2018/03/SAFECode_Fundamental_Practices_for_Secure_Software_Development_March_2018.pdf}},
    year={2018},
    month={03}
    }

@inproceedings {NaiakshinaPasswordStudents2018,
author = {Alena Naiakshina and Anastasia Danilova and Christian Tiefenau and Matthew Smith},
title = {{Deception Task Design in Developer Password Studies: Exploring a Student Sample}},
booktitle = {Fourteenth Symposium on Usable Privacy and Security (SOUPS 2018)},
year = {2018},
isbn = {978-1-939133-10-6},
address = {Baltimore, MD},
pages = {297--313},
url = {https://www.usenix.org/conference/soups2018/presentation/naiakshina},
publisher = {USENIX Association},
month = aug
}

@inproceedings {Palombo2020,
author = {Hernan Palombo and Armin Ziaie Tabari and Daniel Lende and Jay Ligatti and Xinming Ou},
title = {{An Ethnographic Understanding of Software ({In)Security} and a {Co-Creation} Model to Improve Secure Software Development}},
booktitle = {Sixteenth Symposium on Usable Privacy and Security (SOUPS 2020)},
year = {2020},
isbn = {978-1-939133-16-8},
pages = {205--220},
url = {https://www.usenix.org/conference/soups2020/presentation/palombo},
publisher = {USENIX Association},
month = aug
}

@inproceedings{AssalDevSurvey2019,
author = {Assal, Hala and Chiasson, Sonia},
title = {{ 'Think secure from the beginning': A Survey with Software Developers}},
year = {2019},
isbn = {9781450359702},
publisher = {Association for Computing Machinery},
address = {New York, NY, USA},
url = {https://doi.org/10.1145/3290605.3300519},
doi = {10.1145/3290605.3300519},
booktitle = {Proceedings of the 2019 CHI Conference on Human Factors in Computing Systems},
pages = {1–13},
numpages = {13},
location = {Glasgow, Scotland Uk},
series = {CHI '19}
}

@inproceedings{naiakshina2020conducting,
author = {Naiakshina, Alena and Danilova, Anastasia and Gerlitz, Eva and Smith, Matthew},
title = {{On Conducting Security Developer Studies with CS Students: Examining a Password-Storage Study with CS Students, Freelancers, and Company Developers}},
year = {2020},
isbn = {9781450367080},
publisher = {Association for Computing Machinery},
address = {New York, NY, USA},
url = {https://doi.org/10.1145/3313831.3376791},
doi = {10.1145/3313831.3376791},
booktitle = {Proceedings of the 2020 CHI Conference on Human Factors in Computing Systems},
pages = {1–13},
numpages = {13},
location = {Honolulu, HI, USA},
series = {CHI '20}
}

@inproceedings{NaiakshinaPasswordStorage2017,
author = {Naiakshina, Alena and Danilova, Anastasia and Tiefenau, Christian and Herzog, Marco and Dechand, Sergej and Smith, Matthew},
title = {{Why Do Developers Get Password Storage Wrong? A Qualitative Usability Study}},
year = {2017},
isbn = {9781450349468},
publisher = {Association for Computing Machinery},
address = {New York, NY, USA},
url = {https://doi.org/10.1145/3133956.3134082},
doi = {10.1145/3133956.3134082},
booktitle = {Proceedings of the 2017 ACM SIGSAC Conference on Computer and Communications Security},
pages = {311–328},
numpages = {18},
location = {Dallas, Texas, USA},
series = {CCS '17}
}

@inproceedings{FultonBIBIFIWinter2022,
author = {Fulton, Kelsey R. and Votipka, Daniel and Abrokwa, Desiree and Mazurek, Michelle L. and Hicks, Michael and Parker, James},
title = {{Understanding the How and the Why: Exploring Secure Development Practices through a Course Competition}},
year = {2022},
isbn = {9781450394505},
publisher = {Association for Computing Machinery},
address = {New York, NY, USA},
url = {https://doi.org/10.1145/3548606.3560569},
doi = {10.1145/3548606.3560569},
booktitle = {Proceedings of the 2022 ACM SIGSAC Conference on Computer and Communications Security},
pages = {1141–1155},
numpages = {15},
location = {Los Angeles, CA, USA},
series = {CCS '22}
}

@INPROCEEDINGS{acar2016you,
  author={Acar, Yasemin and Backes, Michael and Fahl, Sascha and Kim, Doowon and Mazurek, Michelle L. and Stransky, Christian},
  booktitle={2016 IEEE Symposium on Security and Privacy (SP)}, 
  title={{You Get Where You're Looking for: The Impact of Information Sources on Code Security}}, 
  year={2016},
  volume={},
  number={},
  pages={289-305},
  doi={10.1109/SP.2016.25}}

@inproceedings {binkhorst2022security,
author = {Veroniek Binkhorst and Tobias Fiebig and Katharina Krombholz and Wolter Pieters and Katsiaryna Labunets},
title = {{Security at the End of the Tunnel: The Anatomy of {VPN} Mental Models Among Experts and {Non-Experts} in a Corporate Context}},
booktitle = {31st USENIX Security Symposium (USENIX Security 22)},
year = {2022},
isbn = {978-1-939133-31-1},
address = {Boston, MA},
pages = {3433--3450},
url = {https://www.usenix.org/conference/usenixsecurity22/presentation/binkhorst},
publisher = {USENIX Association},
month = aug
}

@inproceedings {mhaidli2019we,
author = {Abraham H. Mhaidli and Yixin Zou and Florian Schaub},
title = {{"We Can{\textquoteright}t Live Without {Them!}" App Developers{\textquoteright} Adoption of Ad Networks and Their Considerations of Consumer Risks}},
booktitle = {Fifteenth Symposium on Usable Privacy and Security (SOUPS 2019)},
year = {2019},
isbn = {978-1-939133-05-2},
address = {Santa Clara, CA},
pages = {225--244},
url = {https://www.usenix.org/conference/soups2019/presentation/mhaidli},
publisher = {USENIX Association},
month = aug
}

@INPROCEEDINGS{hazhirpasand2021hurdles,
  author={Hazhirpasand, Mohammadreza and Nierstrasz, Oscar and Shabani, Mohammadhossein and Ghafari, Mohammad},
  booktitle={2021 IEEE International Conference on Software Maintenance and Evolution (ICSME)}, 
  title={{Hurdles for Developers in Cryptography}}, 
  year={2021},
  volume={},
  number={},
  pages={659-663},
  doi={10.1109/ICSME52107.2021.00076}}

@INPROCEEDINGS{FischerSOHarmful2017,
  author={Fischer, Felix and Böttinger, Konstantin and Xiao, Huang and Stransky, Christian and Acar, Yasemin and Backes, Michael and Fahl, Sascha},
  booktitle={2017 IEEE Symposium on Security and Privacy (SP)}, 
  title={{Stack Overflow Considered Harmful? The Impact of Copy\&Paste on Android Application Security}}, 
  year={2017},
  volume={},
  number={},
  pages={121-136},
  doi={10.1109/SP.2017.31}
  }

@misc{stackoverflow_survey, title={{Stack Overflow Developer Survey, 2025}}, url={https://survey.stackoverflow.co/2025/ai#sentiment-and-usage-ai-sel-prof-exp}, journal={AI | 2025 Stack Overflow Developer Survey}, publisher={Stack Overflow}, author={Stack Overflow}}

@misc{microsoft_manual_review, title={Security development and operations overview}, howpublished = {\url{{https://www.microsoft.com/en-us/securityengineering/sdl/practices}}}, author={Microsoft}}

@misc{supplemental,
  title = {Supplemental Materials},
  note = {\url{https://osf.io/kdwnp/overview?view_only=34f6ed71dc8f4f9996225f33cbb29612}},
  publisher = {OSF}
}

@inproceedings{bansal-2021-AI-explanations,
author = {Bansal, Gagan and Wu, Tongshuang and Zhou, Joyce and Fok, Raymond and Nushi, Besmira and Kamar, Ece and Ribeiro, Marco Tulio and Weld, Daniel},
title = {{Does the Whole Exceed its Parts? The Effect of AI Explanations on Complementary Team Performance}},
year = {2021},
isbn = {9781450380966},
publisher = {Association for Computing Machinery},
address = {New York, NY, USA},
url = {https://doi.org/10.1145/3411764.3445717},
doi = {10.1145/3411764.3445717},
booktitle = {Proceedings of the 2021 CHI Conference on Human Factors in Computing Systems},
articleno = {81},
numpages = {16},
location = {Yokohama, Japan},
series = {CHI '21}
}

@article{Moore-2008-overestimation,
author = {Moore, Don and Healy, Paul},
year = {2008},
month = {04},
pages = {502-517},
title = {The Trouble With Overconfidence},
volume = {115},
journal = {Psychological Review},
doi = {10.1037/0033-295X.115.2.502}
}

@article{Castelo-2019-aversion,
 ISSN = {00222437, 15477193},
 URL = {https://www.jstor.org/stable/26967271},
 author = {Noah Castelo and Maarten W. Bos and Donald R. Lehmann},
 journal = {Journal of Marketing Research},
 number = {5},
 pages = {pp. 809--825},
 publisher = {[Sage Publications, Inc., American Marketing Association]},
 title = {{Task-Dependent Algorithm Aversion}},
 volume = {56},
 year = {2019}
}

@article{Logg-2019-appreciation,
title = {{Algorithm appreciation: People prefer algorithmic to human judgment}},
journal = {Organizational Behavior and Human Decision Processes},
volume = {151},
pages = {90-103},
year = {2019},
issn = {0749-5978},
doi = {https://doi.org/10.1016/j.obhdp.2018.12.005},
url = {https://www.sciencedirect.com/science/article/pii/S0749597818303388},
author = {Jennifer M. Logg and Julia A. Minson and Don A. Moore}
}

@article{Chaiken-1980-heuristic,
author = {Chaiken, Shelly},
year = {1980},
month = {11},
pages = {752-766},
title = {{Heuristic versus systematic information processing and the use of source versus message cues in persuasion}},
volume = {39},
journal = {Journal of Personality and Social Psychology},
doi = {10.1037/0022-3514.39.5.752}
}

@misc{project_glasswing,
author = {{Anthropic}},
title = {Project Glasswing: An initial update},
year = {2026},
month = {May},
howpublished ={\url{https://www.anthropic.com/research/glasswing-initial-update}},
note = {Accessed: 2026-11-06}
}

@misc{veracode2026,
author = {{Veracode}},
title = {2026 {GenAI} Code Security Report},
year = {2026},
howpublished ={\url{https://www.veracode.com/blog/spring-2026-genai-code-security/}}
}

@INPROCEEDINGS{OmarCodeProp,
  author={Bai, Wei and Akgul, Omer and Mazurek, Michelle L.},
  booktitle={2019 IEEE Cybersecurity Development (SecDev)}, 
  title={{A Qualitative Investigation of Insecure Code Propagation from Online Forums}}, 
  year={2019},
  volume={},
  number={},
  pages={34-48},
  doi={10.1109/SecDev.2019.00016}}

@inproceedings{EdmundsonCodeReview,
author = {Edmundson, Anne and Holtkamp, Brian and Rivera, Emanuel and Finifter, Matthew and Mettler, Adrian and Wagner, David},
title = {{An empirical study on the effectiveness of security code review}},
year = {2013},
isbn = {9783642365621},
publisher = {Springer-Verlag},
address = {Berlin, Heidelberg},
url = {https://doi.org/10.1007/978-3-642-36563-8_14},
doi = {10.1007/978-3-642-36563-8_14},
booktitle = {Proceedings of the 5th International Conference on Engineering Secure Software and Systems},
pages = {197–212},
numpages = {16},
location = {Paris, France},
series = {ESSoS'13}
}

@inproceedings{MeneelyCodeReview,
author = {Meneely, Andrew and Tejeda, Alberto C. Rodriguez and Spates, Brian and Trudeau, Shannon and Neuberger, Danielle and Whitlock, Katherine and Ketant, Christopher and Davis, Kayla},
title = {{An empirical investigation of socio-technical code review metrics and security vulnerabilities}},
year = {2014},
isbn = {9781450332279},
publisher = {Association for Computing Machinery},
address = {New York, NY, USA},
url = {https://doi.org/10.1145/2661685.2661687},
doi = {10.1145/2661685.2661687},
booktitle = {Proceedings of the 6th International Workshop on Social Software Engineering},
pages = {37–44},
numpages = {8},
location = {Hong Kong, China},
series = {SSE 2014}
}

@inproceedings{PaulCodeReview,
author = {Paul, Rajshakhar and Turzo, Asif Kamal and Bosu, Amiangshu},
title = {{Why Security Defects Go Unnoticed during Code Reviews? A Case-Control Study of the Chromium OS Project}},
year = {2021},
isbn = {9781450390859},
publisher = {IEEE Press},
url = {https://doi.org/10.1109/ICSE43902.2021.00124},
doi = {10.1109/ICSE43902.2021.00124},
booktitle = {Proceedings of the 43rd International Conference on Software Engineering},
pages = {1373–1385},
numpages = {13},
location = {Madrid, Spain},
series = {ICSE '21}
}

@inproceedings{BrazCodeReview,
author = {Braz, Larissa and Aeberhard, Christian and \c{C}alikli, G\"{u}l and Bacchelli, Alberto},
title = {{Less is more: supporting developers in vulnerability detection during code review}},
year = {2022},
isbn = {9781450392211},
publisher = {Association for Computing Machinery},
address = {New York, NY, USA},
url = {https://doi.org/10.1145/3510003.3511560},
doi = {10.1145/3510003.3511560},
booktitle = {Proceedings of the 44th International Conference on Software Engineering},
pages = {1317–1329},
numpages = {13},
location = {Pittsburgh, Pennsylvania},
series = {ICSE '22}
}

@inproceedings {AssalSDLC18,
author = {Hala Assal and Sonia Chiasson},
title = {Security in the Software Development Lifecycle},
booktitle = {Fourteenth Symposium on Usable Privacy and Security (SOUPS 2018)},
year = {2018},
isbn = {978-1-939133-10-6},
address = {Baltimore, MD},
pages = {281--296},
url = {https://www.usenix.org/conference/soups2018/presentation/assal},
publisher = {USENIX Association},
month = aug
}

@misc{AmazonKiro,
  title = {{Amazon Kiro}},
  note = {\url{https://kiro.dev}},
  publisher = {Amazon}
}

@misc{CursorPlanMode,
  title = {{Cursor Plan Mode}},
  note = {\url{https://cursor.com/docs/agent/plan-mode}},
  publisher = {Cursoe}
}

\appendices
\section{}
\subsection{Open Science}
\label{app:openscience}
In the spirit of transparency and reproducibility, we provide the following artifacts in our supplemental materials~\cite{supplemental}: (1) The surveys used for screening and post-coding exercise, (2) the instructions, tasks, and code snippets provided to participants when using the NERDS system for the coding tasks,  and (3) the anonymized data used for producing the models in Tables~\ref{tab:num-vulns},~\ref{tab:models_combo},~\ref{tab:rq3model},~\ref{tab:usefulness-of-AI-regression},~\ref{tab:future-AI-use-regression}.
Our appendix and supplementary materials~\cite{supplemental} contain additional information related to the study, including the post-exercise survey, the interview script, and additional data not included in the main paper.

We do not release raw interview recordings or full transcripts to protect participants' privacy and confidentiality. This is in accordance with our institution's ethics review board's policies and ethical research practices for similar studies. As stated in our institution's ethics review board approved protocol, all recordings were permanently deleted immediately after we validated the transcripts for accuracy. We present our findings through thematic analysis with anonymized quotes.

\subsection{Additional Quantitative Data}
\label{sec:appendix}
To provide additional context for our results, this appendix includes a more thorough breakdown of the sampled population along with the CWEs and functionality we tested for each of the 400 submissions. Our supplementary materials~\cite{supplemental} contain the full set of unique vulnerabilities we considered while evaluating participants' submissions. Table \ref{tab:demographics} presents the demographics for all 100 study participants, and Table \ref{tab:interview-demographics} presents the demographics for the 23 participants who were interviewed.

\begin{table}[H]
    \centering
    \footnotesize
    \begin{tabular}{l c c c}
        \toprule
        \textbf{Task Name} & \textbf{\% Secure} & \textbf{Avg. \# of vulns.} \\
         & \textbf{Submissions} & \\
        \midrule
        Add Item     & 8\% & 2.9 \\
        Update Item  & 25\% & 1.9 \\
        Swap Item    & 26\% & 2.1 \\
        Remove Item  & 29\% & 1.6 \\
        \midrule
        \textbf{Total} & \textbf{22\%} & \textbf{2.1}  \\
        \bottomrule
    \end{tabular}
    \caption{Final Solution Security by Task.}
    \label{tab:solution-security}
\end{table}

\begin{table}[]
    \centering
    \begin{threeparttable}
    \footnotesize
    \begin{tabular}{lccc}
        \toprule
        \textbf{PID} & \textbf{Prog. Exp.} & \textbf{Sec. Exp.} \\ %
        \midrule
        P1 & 6 & 1 \\%
        P2 & 13 & 2 \\%
        P3 & 6 & 1 \\%
        P4 & 4 & 0 \\%
        P5 & 3 & 1 \\%
        P6 & 4 & 2 \\%
        P7 & 5 & 1 \\%
        P8 & 5 & 0 \\%
        P9 & 20 & 2 \\%
        P10 & 4 & 0.5 \\%
        P11 & 10 & 1 \\%
        P12 & 3.5 & 1 \\%
        P13 & 5 & 1 \\
        P14 & 4 & 0 \\%
        P15 & 14 & 0.5 \\%
        P16 & 8 & 1 \\%
        P17 & 3 & 0.5 \\%
        P18 & 10 & 0 \\%
        P19 & 4 & 0 \\%
        P20 & 3 & 2 \\%
        P21 & 5 & 2 \\%
        P22 & 6 & 0 \\%
        P23 & 20 & 5 \\%
        \bottomrule
    \end{tabular}
    \end{threeparttable}
    \caption{Interview participants' experience and vulnerability statistics. \textbf{PID} shows the participant ID. \textbf{Prog. Exp.} and \textbf{Sec. Exp.} denote years of programming and security experience, respectively.}

    \label{tab:interview-demographics}
\end{table}

\begin{table}[t]
\centering
\footnotesize
\begin{tabular}{llrrr}
\toprule
\textbf{Variable} & \textbf{Value}  & \textbf{OR} & \textbf{CI} & \textbf{\textit{p}-value} \\
\midrule
prog. exp.  &  $\le3 years$ & --  \\
 & $3+$ years & 2.67 & [0.72, 9.89] & 0.142  \\
\bottomrule
\end{tabular}
\caption{Logistic regression model to see what impacts participants' perceptions of the usefulness of AI.}
\label{tab:usefulness-of-AI-regression}
\end{table}

\begin{table}[t]
\centering
\begin{tabular}{@{}llrrr@{}}
\toprule
\textbf{Grouping} & \textbf{Level} & \textbf{Functional} & \textbf{Secure} & \textbf{Frac. secure} \\
\midrule
\multirow{4}{*}{Task}
& add item & 96 & 8 & 0.083 \\
& remove item & 71 & 29 & 0.408 \\
& swap item & 93 & 26 & 0.280 \\
& update item & 95 & 25 & 0.263 \\
\midrule
\multirow{2}{*}{\shortstack[l]{Security\\experience}}
& $\leq$3 years & 301 & 69 & 0.229 \\
& 3+ years & 54 & 19 & 0.352 \\
\midrule
\multirow{2}{*}{\shortstack[l]{Programming\\experience}}
& $\leq$3 years & 84 & 30 & 0.357 \\
& 3+ years & 271 & 58 & 0.214 \\
\midrule
\multirow{5}{*}{\shortstack[l]{Snippet\\chosen}}
& 1 & 100 & 23 & 0.230 \\
& 2 & 91 & 15 & 0.165 \\
& 3 & 86 & 22 & 0.256 \\
& 4 & 41 & 19 & 0.463 \\
& 5 & 37 & 9 & 0.243 \\
\midrule
\multirow{4}{*}{\shortstack[l]{Task\\number}}
& 1 & 90 & 18 & 0.200 \\
& 2 & 87 & 21 & 0.241 \\
& 3 & 90 & 25 & 0.278 \\
& 4 & 88 & 24 & 0.273 \\
\midrule
\multirow{4}{*}{\shortstack[l]{Difference in \\LOC (factor of \\original)}}
& [0.00, 0.11] & 86 & 0 & 0.000 \\
& (0.11, 0.95] & 77 & 5 & 0.065 \\
& (0.95, 1.78] & 94 & 32 & 0.340 \\
& (1.78, 4.86] & 98 & 51 & 0.520 \\
\midrule
\multirow{4}{*}{\shortstack[l]{Number of\\external links\\visited}}
& 0 & 172 & 52 & 0.302 \\
& 1 & 53 & 4 & 0.076 \\
& 2 & 42 & 10 & 0.238 \\
& 3+ & 88 & 22 & 0.250 \\
\bottomrule \\
\end{tabular}
\caption{Fraction of secure solutions by task and participant characteristics.}
\label{tab:secure-frac}
\end{table}

\begin{table}[t]
\centering
\footnotesize
\begin{tabular}{llrrr}
\toprule
\textbf{Variable} & \textbf{Value}  & \textbf{OR} & \textbf{CI} & \textbf{\textit{p}-value} \\
\midrule

prog. exp.  &  $\le3 years$  & --  \\
 & $3 +$ years &  0.30 & [0.04, 2.52] & 0.270  \\

\bottomrule
\end{tabular}
\caption{Logistic regression model to see what impacts participants' desire to use AI in the future.}
\label{tab:future-AI-use-regression}
\end{table}

\begin{table}[]
\footnotesize
\begin{tabular}{rlc}
\toprule
\textbf{Demographic} & \textbf{Value} & \textbf{N} \\ \midrule
Gender & Male & 91 \\
& Female & 8 \\
& Prefer Not to Respond & 1 \\ \midrule
Age &    18-29 & 83 \\
(years)& 30-39 & 9 \\
& 40-49 & 2 \\
& 50-59 & 4 \\
& 60-69 & 1 \\
& Prefer Not to Respond & 1 \\ \midrule
Ethnicity  & Asian & 51 \\
& White & 26 \\
& Hispanic/Latino & 4 \\
& Black/African American & 4 \\
& Other & 6 \\
& Prefer Not to Respond & 9 \\ \midrule
Country of Residence & Pakistan & 24 \\
& US & 12 \\
& India & 12\\
& Egypt & 7 \\
& Turkey & 4 \\
& Other & 41 \\ \midrule
Education & Bachelor's &  48 \\
&  Master's & 19 \\
 & Some college no degree & 16\\
& High School diploma & 9  \\
& Other &  8       \\\midrule
Field of study           &  Computer Science & 73 \\
& Other Engineering & 16 \\
& Cybersecurity/ IT security & 2 \\
& Other & 8 \\
& Prefer Not to Respond & 1 \\\midrule
Security experience      & Min & 0 \\
(years)& Max & 20 \\
& Mean & 1.8 \\
& Median & 1 \\
& Std. dev. & 3.0 \\\midrule
Programming experience   & Min & 1 \\
(years)& Max &  40 \\
& Mean & 7.0 \\
& Median & 5 \\
& Std. dev. & 6.0 \\\midrule
C programming experience & Min & 0 \\
(years)& Max &  37 \\
& Mean &  4.4 \\
& Median &  3 \\
& Std. dev. &  4.8 \\\bottomrule
\end{tabular}
\caption{Demographics for 100 participants.}
\label{tab:demographics}
\end{table}

\subsection{Ethical Considerations}
\label{app:ethics}
Our study was reviewed and approved as an exempt research study by the primary author's ethics review board. We ensured that the study was conducted in accordance with the Menlo Report standards and was GDPR compliant.

Informed consent was obtained from participants before any data collection began. Additionally, participants residing in a country covered by the GDPR were asked to provide consent for data collection. Participants were also reminded of the consent document before beginning the task on the NERDS system if they qualified from the pre-screening survey. Finally, participants selected for interviews were asked to reaffirm their consent before recording began. Throughout, we made it clear that participation in the study was voluntary and that participants could withdraw at any time without penalty. We anonymized all collected data to protect participants' privacy and confidentiality.

We compensated all participants for their time and expertise. All participants were given a \$10 gift card (\$60/hour if they only completed the screening survey); those who were qualified for the full study and completed all four tasks were given an additional \$20 regardless of their performance, giving a total compensation of \$30 (\$20/hour). Participants who completed all four tasks and either had perfect scores in functionality and security or were in the top 10\% of all responses were given an additional \$20, for a total of \$50. Finally, any participants who were selected and took part in the interviews received an additional \$20, which means these participants would be compensated either \$50 or \$70, depending on their performance in the coding exercise.

The only potential harm we envision from publishing this work is that malicious actors may target AI systems to induce them to produce more insecure code, given that many participants did not validate the security of their code or seek external security resources. However, the risks associated with this are relatively low, given that prior work has established that AI systems already produce insecure code~\cite{PerryInsecureAICode,HamerAnotherCopyPaste2024,PearceAsleep2022}. Conversely, we believe that our work is important for the security community, AI system developers, and developers writ large. We show that developers do not audit for security, instead prioritizing functionality, and that the security of their code is related to how secure the initial suggestion is; therefore, AI systems that generate code must prioritize highlighting security.

\end{document}